\documentclass[amsmath,amssymb]{revtex4}
\usepackage{graphicx}% Include figure files
\usepackage{amscd,amsmath,amsthm,amsfonts,amssymb}
\usepackage{ytableau}
\usepackage{extarrows}
\def\[{\begin{equation}}
\def\]{\end{equation}}
\newtheorem{lem}{Lemma}
\newtheorem{thm}{Theorem}
\begin{document}
\title{Triangular rogue clusters associated with multiple roots of Adler--Moser polynomials in integrable systems}
\author{%%%% Author details
Bo Yang$^{1}$, Jianke Yang$^{2}$}
%%%%%%%%% Insert author address here
\address{$^{1}$ School of Mathematics and Statistics, Ningbo University, Ningbo 315211, China\\
$^{2}$ Department of Mathematics and Statistics, University of Vermont, Burlington, VT 05401, U.S.A}
\begin{abstract}
Rogue patterns associated with multiple roots of Adler--Moser polynomials under general multiple large parameters are studied in integrable systems. It is first shown that the multiplicity of any multiple root in any Adler--Moser polynomial is a triangular number (i.e., its multiplicity is equal to $n(n+1)/2$ for a certain integer $n$). Then, it is shown that corresponding to a nonzero multiple root of the Adler--Moser polynomial, a triangular rogue cluster would appear on the spatial-temporal plane. This triangular rogue cluster comprises $n(n+1)/2$ fundamental rogue waves forming a triangular shape, and space-time locations of fundamental rogue waves in this triangle are a linear transformation of the Yablonskii--Vorob'ev polynomial $Q_{n}(z)$'s root structure. In the special case where this multiple root of the Adler--Moser polynomial is zero, the associated rogue pattern is found to be a $n$-th order rogue wave in the $O(1)$ neighborhood of the spatial-temporal origin. These general results are demonstrated on two integrable systems: the nonlinear Schr\"odinger equation and the generalized derivative nonlinear Schr\"odinger equation. For these equations, asymptotic predictions of rogue patterns are compared with true rogue solutions and good agreement between them is illustrated.
\end{abstract}
\maketitle
\section{Introduction}

Rogue waves, also known as freak waves, monster waves and extreme waves, are unusually large and suddenly appearing surface waves in the sea \cite{Ocean_rogue_review,Pelinovsky_book}. Since they appear and disappear without warning, they can be dangerous to ships, even to large ones. In order to understand the mathematical and physical mechanisms of these waves, an important theoretical discovery was that the nonlinear Schr\"{o}dinger (NLS) equation that governs one-dimensional wave-packet propagation in the ocean admits rational solutions that show rogue-like behaviors \cite{Peregrine1983,AAS2009,DGKM2010,Akhmediev_triplet2011,KAAN2011,GLML2012,OhtaJY2012}. Since the NLS equation also governs wave propagation in many other physical systems such as optics and plasma, this implies that rogue waves could appear in those other physical systems as well. Such predictions were subsequently verified in experiments of optics, water waves and plasma \cite{Fiber1,Tank1,Tank2,Plasma,Chabchoub_triplet2013}, which significantly deepened our understanding of physical rogue events. Due to this success, rogue wave solutions in many other integrable equations have also been derived, and some of those solutions have been observed in experiments as well (see \cite{Yangbook2024} for a review).

Pattern formation of rogue waves is an important issue as this information could allow for the prediction of later rogue wave events from earlier wave forms. In the NLS equation, some interesting patterns of rogue solutions were numerically plotted in \cite{KAAN2013}, but this numerical plotting quickly became difficult as the order of the solution increased. A new discovery we made in the past few years was that, clear rogue patterns in the NLS equation would appear when internal parameters in its rogue wave solutions get large, and such rogue patterns could be predicted asymptotically by the root structures of certain special polynomials \cite{Yang2021a,YangAD2024}. If a single internal parameter is large, the rogue pattern would be predicted by the  root structure of a certain Yablonskii--Vorob'ev hierarchy polynomial, with each simple root inducing a fundamental rogue wave in the spatial-temporal plane and a multiple zero root inducing a lower-order rogue wave in the neighborhood of the spatial-temporal origin \cite{Yang2021a}. If multiple internal parameters are large in a single-power form, the rogue pattern would be predicted by the root structure of a certain Adler--Moser polynomial, with each simple root inducing a fundamental rogue wave in the spatial-temporal plane \cite{YangAD2024}. The case of wave patterns induced by a multiple root in the corresponding Adler--Moser polynomial was considered recently in \cite{Ling2025}. It was shown that for a more involved form of multiple large parameters with special coefficient values, a multiple root in the Adler--Moser polynomial could induce various rogue shapes. Their special choices of multiple large parameters do not contain the multiple large parameters of single-power form we considered in \cite{YangAD2024}, however.

In this paper, we study rogue patterns associated with multiple roots of Adler--Moser polynomials under general multiple large parameters of single-power form in integrable systems and complete the work we started in \cite{YangAD2024}. We first show that the multiplicity of any multiple root in any Adler--Moser polynomial is a triangular number (i.e., its multiplicity is equal to $n(n+1)/2$ for a certain integer $n$). We then show that corresponding to a nonzero multiple root of the Adler--Moser polynomial, a triangular rogue cluster would appear on the spatial-temporal plane. This triangular rogue cluster comprises $n(n+1)/2$ fundamental rogue waves forming a triangular shape, and space-time locations of fundamental rogue waves in this triangle are a linear transformation of the Yablonskii--Vorob'ev polynomial $Q_{n}(z)$'s root structure. In the special case where this multiple root of the Adler--Moser polynomial is zero, we show that the associated rogue pattern is a $n$-th order rogue wave in the $O(1)$ neighborhood of the spatial-temporal origin. We demonstrate these general results on two integrable systems: the NLS equation and the generalized derivative nonlinear Schr\"odinger (GDNLS) equations. For these equations, we compare our asymptotic predictions of rogue patterns with true rogue solutions and show good agreement between them.

\section{Preliminaries}

First, we introduce Schur polynomials $S_j(\mbox{\boldmath $x$})$, where $\emph{\textbf{x}}=\left( x_{1}, x_{2}, \ldots \right)$. These polynomials are defined by
\begin{equation} \label{Schurdef}
\sum_{j=0}^{\infty}S_j(\mbox{\boldmath $x$}) \epsilon^j
=\exp\left(\sum_{i=1}^{\infty}x_i \epsilon^i\right),
\end{equation}
or more explicitly,
\begin{equation} \label{Skdef}
S_{j}(\mbox{\boldmath $x$}) =\sum_{l_{1}+2l_{2}+\cdots+ml_{m}=j} \left( \ \prod _{i=1}^{m} \frac{x_{i}^{l_{i}}}{l_{i}!}\right).
\end{equation}
In particular, $S_0(\mbox{\boldmath $x$})=0$ and $S_1(\mbox{\boldmath $x$})=x_1$. We also define $S_j(\mbox{\boldmath $x$})\equiv 0$ when $j<0$.

Next, we introduce two types of special polynomials that will be important for our work.

\subsection{Yablonskii--Vorob'ev polynomials and their root structures}

Yablonskii--Vorob'ev polynomials arose in rational solutions of the second Painlev\'{e} equation ($\mbox{P}_{\mbox{\scriptsize II}}$) \cite{Yablonskii1959,Vorobev1965}.
\[\label{PII}
w''=2 w^3+ z w+\alpha,
\]
where the prime denotes derivative to the variable $z$, and $\alpha$ is an arbitrary constant. It has been shown that this $\mbox{P}_{\mbox{\scriptsize II}}$ equation admits rational solutions if and only if $\alpha=N$ is an integer. In this case, the rational solution is unique and is given by
\begin{eqnarray}
&& w(z; N)= \frac{d}{dz} \ln\frac{Q_{N-1}(z)}{Q_{N}(z)}, \quad N\ge 1,   \label{wzn1}\\
&& w(z; 0)=0, \quad w(z; -N)=-w(z; N),      \label{wzn2}
\end{eqnarray}
and the polynomials $Q_N(z)$, now called the Yablonskii--Vorob'ev polynomials, are constructed by the following recurrence relation
\[
Q_{N+1}Q_{N-1}= z Q_{N}^2 -4 \left[ Q_{N}Q_{N}''-(Q_{N}')^2 \right],
\]
with $Q_{0}(z)=1$, $Q_{1}(z)=z$, and the prime denoting the derivative to $z$. Later, a determinant expression for these polynomials was found in \cite{Kajiwara-Ohta1996}. Let $p_{j}(z)$ be the special Schur polynomial defined by
\begin{equation} \label{defpk}
\sum_{j=0}^{\infty}p_j(z) \epsilon^j =\exp\left( z \epsilon - \frac{4}{3}\epsilon^3 \right),
\end{equation}
and $p_{j}(z)\equiv 0$ if $j<0$. Then, Yablonskii--Vorob'ev polynomials $Q_{N}(z)$ are given by the $N \times N$ determinant \cite{Kajiwara-Ohta1996}
\begin{eqnarray}
&& Q_{N}(z) = c_{N} \left| \begin{array}{cccc}
         p_{1}(z) & p_{0}(z) & \cdots &  p_{2-N}(z) \\
         p_{3}(z) & p_{2}(z) & \cdots &  p_{4-N}(z) \\
        \vdots& \vdots & \vdots & \vdots \\
         p_{2N-1}(z) & p_{2N-2}(z) & \cdots &  p_{N}(z)
       \end{array}
 \right|,
\end{eqnarray}
where $c_{N}= \prod_{j=1}^{N}(2j-1)!!$. This determinant is a Wronskian since we can see from Eq.~(\ref{defpk}) that $p'_{j}(z)=p_{j-1}(z)$. The $Q_{N}(z)$ polynomial is monic with integer coefficients and has degree $N(N+1)/2$ \cite{Clarkson2003-II}. The first few such polynomials are
\begin{eqnarray*}
&& Q_2(z)=z^3+4, \\
&& Q_3(z)=z^6 + 20z^3 - 80,  \\
&& Q_4(z)=z(z^9 + 60z^6 + 11200).
\end{eqnarray*}

\begin{figure}[htb]
\begin{center}
\includegraphics[scale=0.30, bb=650 0 800 400]{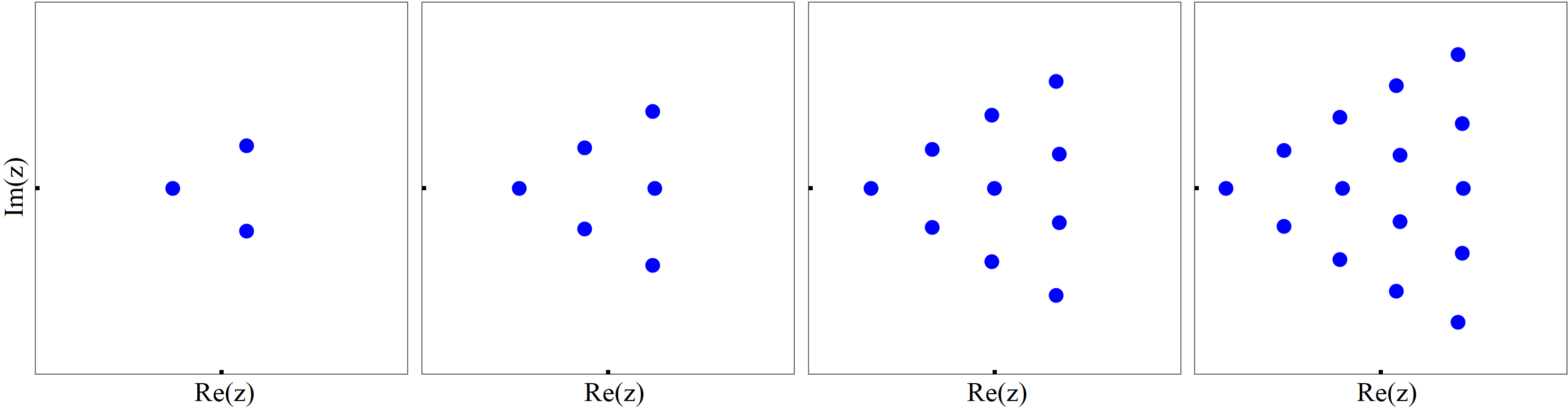}
\caption{Root structures of Yablonskii--Vorob'ev polynomials $Q_{N}(z)$ for $N=2, 3, 4, 5$ (from left to right). In all panels, $ -6 \leq \Re(z),\Im(z) \leq 6$, where $\Re$ and $\Im$ represent the real and imaginary parts of a complex number respectively. } \label{f:rootYV}
\end{center}
\end{figure}

Root structures of Yablonskii--Vorob'ev polynomials will be important to us. It was shown in \cite{Fukutani} that all roots of $Q_{N}(z)$ are simple. It was further shown in \cite{Clarkson2003-II,Miller2014,Bertola2016} that these simple roots form a triangular shape (the three edges of this triangular shape are not completely straight; but we will still call it a triangle for simplicity). To illustrate, we display these triangular root patterns of $Q_{N}(z)$ for $2\le N\le 5$ in Fig.~\ref{f:rootYV}.

\subsection{Adler--Moser polynomials}

Adler--Moser polynomials were proposed by Adler and Moser \cite{Adler_Moser1978}, who expressed rational solutions of the
Korteweg-de Vries equation in terms of those polynomials. In a different context of point vortex dynamics, it was discovered unexpectedly that the zeros of these polynomials also form stationary vortex configurations when the vortices have the same strength but positive or negative orientations \cite{Aref2007FDR,Clarkson2009}.

Adler--Moser polynomials $\Theta_{N}(z)$ can be written as a determinant \cite{Clarkson2009}
\begin{equation} \label{AdlerMoserdef}
\Theta_{N}(z) = c_{N} \left| \begin{array}{cccc}
         \theta_{1}(z) &\theta_{0}(z) & \cdots &  \theta_{2-N}(z) \\
         \theta_{3}(z) & \theta_{2}(z) & \cdots &  \theta_{4-N}(z) \\
        \vdots& \vdots & \vdots & \vdots \\
         \theta_{2N-1}(z) & \theta_{2N-2}(z) & \cdots &  \theta_{N}(z)
       \end{array} \right|,
\end{equation}
where $\theta_{j}(z)$ are Schur polynomials defined by
\begin{equation}\label{AMthetak}
\sum_{j=0}^{\infty} \theta_j(z) \epsilon^j =\exp\left( z \epsilon + \sum_{i=1}^{\infty} \kappa_i \epsilon^{2i+1} \right),
\end{equation}
$\theta_{j}(z)\equiv 0$ if $j<0$, and $\kappa_j \hspace{0.04cm} (j\ge 1)$ are arbitrary complex constants. Note that our $\kappa_j$ constant is slightly different from that in \cite{Clarkson2009} by a factor of $-1/(2j+1)$, and this different parameter definition will be more convenient for our purpose. The determinant in (\ref{AdlerMoserdef}) is a Wronskian since we can see from Eq.~(\ref{AMthetak}) that $\theta'_{j}(z)=\theta_{j-1}(z)$. In addition, the $\Theta_{N}(z)$ polynomial is monic with degree $N(N+1)/2$, which can be seen by noticing that the highest $z$ term of $\theta_j(z)$ is $z^j/j!$, and the determinant in (\ref{AdlerMoserdef}) with $\theta_j(z)$ replaced by its highest $z$ term can be explicitly calculated as $z^{N(N+1)/2}$ \cite{OhtaJY2012}. Adler--Moser polynomials $\Theta_{N}(z)$ reduce to Yablonskii--Vorob'ev polynomials $Q_{N}(z)$ when we set $\kappa_1=-4/3$ and the other $\kappa_j$ zero. Thus, Adler--Moser polynomials can be viewed as generalizations of Yablonskii--Vorob'ev polynomials.

The first few Adler--Moser polynomials are
\begin{eqnarray*}
&& \hspace{-0.8cm} \Theta_2(z; \kappa_1)=z^3-3 \kappa _1, \\
&& \hspace{-0.8cm} \Theta_3(z; \kappa_1, \kappa_2)=z^6-15 \kappa _1 z^3+45 \kappa _2 z-45 \kappa _1^2,  \\
&& \hspace{-0.8cm} \Theta_4(z; \kappa_1, \kappa_2, \kappa_3)=z^{10}-45 \kappa _1 z^7+315 \kappa _2 z^5-1575 \kappa _3 z^3  +4725 \kappa _1 \kappa _2 z^2-4725 \kappa _1^3 z +4725 (\kappa _1 \kappa _3-\kappa _2^2), \quad \mbox{\hspace{1cm}} \\
&& \hspace{-0.8cm} \Theta_5(z; \kappa_1, \kappa_2, \kappa_3, \kappa_4)=z^{15}-105\kappa_1 z^{12}+1260 \kappa_2 z^{10} +1575\kappa_1^2z^9 -14175\kappa_3 z^8 +14175 \kappa_1\kappa_2 z^7 \\
&&\hspace{2.1cm}  -33075 (\kappa_1^3-3\kappa_4)z^6 -297675(\kappa_2^2+\kappa_1\kappa_3)z^5+1488375\kappa_1^2\kappa_2 z^4 \\
&&\hspace{2.1cm}  -496125 (2\kappa_1^4-3\kappa_2\kappa_3+3\kappa_1\kappa_4)z^3
+4465125\kappa_1(\kappa_1\kappa_3-\kappa_2^2)z^2  \\
&&\hspace{2.1cm}-1488375(\kappa_1^3\kappa_2+3\kappa_3^2-3\kappa_2\kappa_4)z
+1488375(\kappa_1^5-3\kappa_2^3+6\kappa_1\kappa_2\kappa_3-3\kappa_1^2\kappa_4).
\end{eqnarray*}

\section{Multiplicity of multiple roots in Adler--Moser polynomials}

Root structures of Adler--Moser polynomials are important for rogue patterns when the underlying rogue wave possesses multiple large internal parameters. Indeed, we have shown in \cite{YangAD2024,Yangbook2024} that every simple root of the underlying Adler--Moser polynomial gives rise to a fundamental rogue wave whose space-time location is linearly dependent on the value of this simple root. Thus, our focus in this paper will be on multiple roots of Adler--Moser polynomials and the rogue patterns they create. Since Adler--Moser polynomials contain free complex parameters $\{\kappa_j\}$, by choosing those parameters judiciously, multiple roots clearly can be created in these polynomials. We give three examples below.

\begin{figure}[htb]
\begin{center}
\vspace{3.6cm}
\includegraphics[scale=0.35, bb=550 0 600 145]{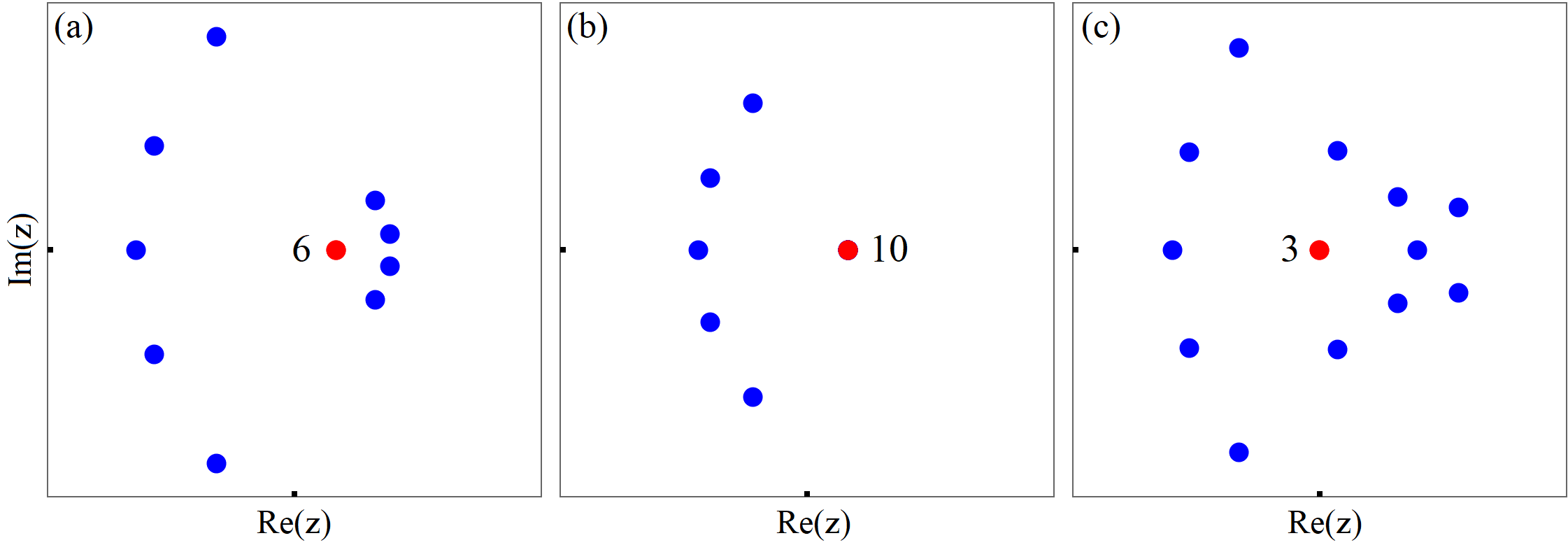}
\caption{Roots of Adler--Moser polynomials $\Theta_{5}(z;\kappa_1, \kappa_2, \kappa_3, \kappa_4)$. The parameter values of $(\kappa_1, \kappa_2, \kappa_3, \kappa_4)$ for panels (a)-(c) are given in Eqs.~(\ref{parameter1})-(\ref{parameter3}) respectively. In all plots, a multiple root is marked by a red dot with its multiplicity indicated by a number beside it. Simple roots are marked by blue dots. In all panels, $-6 \leq \Re(z),\Im(z) \leq 6$. } \label{f:rootAM}
\end{center}
\end{figure}

Our three examples are $\Theta_5(z; \kappa_1, \kappa_2, \kappa_3, \kappa_4)$ polynomials with the following three sets of parameter values,
\[\label{parameter1}
(\kappa_1, \kappa_2, \kappa_3, \kappa_4 ) = \left(1, \frac{59}{45}, \frac{377}{189}, \frac{89}{27}\right),
\]
\[\label{parameter2}
(\kappa_1, \kappa_2, \kappa_3, \kappa_4 ) = \left(\frac{1}{3}, \frac{1}{5}, \frac{1}{7}, \frac{1}{9}\right),
\]
and
\[\label{parameter3}
(\kappa_1, \kappa_2, \kappa_3, \kappa_4 ) = \left(1, 1, 1, \frac{4}{3}\right).
\]
Root structures for these three $\Theta_5(z; \kappa_1, \kappa_2, \kappa_3, \kappa_4)$ polynomials are displayed in Fig.~\ref{f:rootAM}(a, b, c) respectively. It is seen that in the first case (\ref{parameter1}), this polynomial has a root $z=1$ of multiplicity 6, plus 9 other simple roots which form two opposing arcs on the two sides of the $z=1$ root in the complex plane. In the second case (\ref{parameter2}), this polynomial has a root $z=1$ of multiplicity 10, plus 5 other simple roots which form an arc on the left side of the $z=1$ root. This set of parameter values and its root properties have been considered in \cite{Ling2025}. In the third case (\ref{parameter3}), this polynomial has a zero root of multiplicity 3, plus 12 other simple roots which form a complex shape surrounding the zero root. In all three examples, a multiple root appears, and this multiple root is nonzero in the first two cases and zero in the last case.

One may notice that the multiplicities of the multiple roots in these three examples are 3, 6 and 10, which are all triangular numbers, i.e., numbers of the form $j(j+1)/2$ for a certain integer $j$. This is not an accident. Indeed, we will show that the multiplicity of every multiple root in any Adler--Moser polynomial is a triangular number. This result is presented in the following theorem.

\begin{thm}
The multiplicity of every multiple root in any Adler--Moser polynomial is a triangular number.
\end{thm}

This result is important for the prediction of rogue patterns associated with Adler--Moser polynomials, as we will see in later sections. Its proof is given below.

\subsection{Row echelon form of a special matrix} \label{secRowEch}
To prove Theorem 1, we first introduce a lemma on the row echelon form of a special $N\times 2N$ matrix
\[ \label{defHN}
\mathbf{H}=\left(\begin{array}{ccccccccc} h_1 & 1 & & &&&    \\
                               h_3 & h_2 & h_1 & 1 &&&   \\
                               h_5 & h_4 & h_3 & h_2 & h_1 & 1 & \\
                               \vdots & \vdots & \vdots & \vdots & \vdots & \vdots & \\
                               h_{2N-1} & h_{2N-2} & h_{2N-3} &h_{2N-4} & h_{2N-5} & h_{2N-6} & \cdots & h_1 & 1
                               \end{array} \right),
\]
where $h_1, h_2, \dots, h_{2N-1}$ are complex constants. Unwritten elements in this matrix are all zero.

A matrix is in row echelon form if
\begin{enumerate}
\item All rows having only zero entries are at the bottom;
\item The leading entry (that is, the left-most nonzero entry) of every nonzero row is on the right of the leading entry of every row above.
\end{enumerate}
These two conditions imply that all entries in a column below a leading entry are zeros.

Every matrix $\mathbf{H}$ can be reduced to a row echelon form $\widehat{\mathbf{H}}$ through two types of elementary row operations,
\begin{enumerate}
\item interchange two rows;
\item multiply a row by a nonzero number and then add it to a lower row.
\end{enumerate}
The process to reduce $\mathbf{H}$ to its row echelon form $\widehat{\mathbf{H}}$ is called Gauss elimination. In matrix notations, $\mathbf{H}$ and its row echelon form $\widehat{\mathbf{H}}$ are related as
\[ \label{PHLH}
\mathbf{P}\mathbf{H}=\mathbf{L}\widehat{\mathbf{H}},
\]
where $\mathbf{P}$ is a permutation matrix which records type-i row operations, and $\mathbf{L}$ is a lower triangular matrix with ones on the diagonal, which records type-ii row operations.

It turns out that the row echelon form $\widehat{\mathbf{H}}$ of the special matrix $\mathbf{H}$ in Eq.~(\ref{defHN}) has a special structure, and this special structure is presented in the following lemma.

\begin{lem}
The row echelon form $\widehat{\mathbf{H}}$ of the $N\times 2N$ matrix $\mathbf{H}$ in Eq.~(\ref{defHN}) has the following special structure
\[ \label{defHhatN}
\widehat{\mathbf{H}}=\left(\begin{array}{cc} \mathbf{A} & \mathbf{C}    \\  \mathbf{0} & \mathbf{B}                               \end{array} \right),
\]
where $\mathbf{A}$ is a $k\times k$ upper triangular matrix with nonzero diagonal elements, $k$ is the number of the first consecutive columns of $\mathbf{H}$ that are linearly independent, $\mathbf{B}$ is a $(N-k)\times (2N-k)$ matrix of the following staired form
\[ \label{BformLemma1}
\mathbf{B}=\left(\begin{array}{cccccccccccc} 0 & \beta & \dots & \dots &  \dots & \dots & \dots & \dots & \dots & \dots & \dots  \\
                               && 0 & \beta & \dots & \dots & \dots & \dots & \dots & \dots & \dots \\
                               &&&& 0 & \beta & \dots & \dots & \dots  & \dots & \dots \\
                               &&&&&&\ddots & \ddots &\ddots & \ddots &\vdots \\
                               &&&&&&& & 0 & \beta & \dots
                               \end{array} \right),
\]
i.e., the $i$-th row of matrix $\mathbf{B}$ starts with $2i-1$ zeros, followed by $\beta$ and then other row elements, and $\beta\ne 0$.
\end{lem}
Proof of this lemma will be provided in the appendix.

\subsection{Proof of Theorem 1}
Now, we are ready to prove Theorem 1.

Suppose $z_0$ is a multiple root of the Adler--Moser polynomial $\Theta_{N}(z)$. Let us denote
\[ \label{zz0}
z=z_0+\hat{z}, \quad \Theta_{N}(z)=\widehat{\Theta}_{N}(\hat{z}).
\]
Then, $\hat{z}=0$ is a multiple root of the polynomial $\widehat{\Theta}_{N}(\hat{z})$.
When $z=z_0+\hat{z}$ is substituted into Eq.~(\ref{AMthetak}), we get
\begin{equation}\label{AMthetak2}
\sum_{j=0}^{\infty} \theta_j(z) \epsilon^j =e^{\hat{z}\epsilon} \exp\left( z_0 \epsilon + \sum_{i=1}^{\infty} \kappa_i \epsilon^{2i+1} \right).
\end{equation}
From Eq.~(\ref{AMthetak}), we see that
\[ \label{defhn}
\exp\left( z_0 \epsilon + \sum_{i=1}^{\infty} \kappa_i \epsilon^{2i+1} \right)=\sum_{j=0}^\infty h_j \epsilon^j,
\]
where $h_j=\theta_j(z_0)$. Using this expansion and the Taylor expansion of $e^{\hat{z}\epsilon}$, we get from Eq.~(\ref{AMthetak2}) that
\[
\theta_j(z)=\sum_{i=0}^j h_{j-i} \frac{\hat{z}^i}{i!}.
\]
Thus,
\[ \label{thetaH}
\left( \begin{array}{cccc}
         \theta_{1}(z) &\theta_{0}(z) & \cdots &  \theta_{2-N}(z) \\
         \theta_{3}(z) & \theta_{2}(z) & \cdots &  \theta_{4-N}(z) \\
        \vdots& \vdots & \vdots & \vdots \\
         \theta_{2N-1}(z) & \theta_{2N-2}(z) & \cdots &  \theta_{N}(z)
       \end{array} \right)_{N\times N}=\mathbf{H} \left( \begin{array}{cccc}
         1 &0 & \cdots &  0 \\
         \hat{z} & 1 & \cdots &  0 \\
        \vdots& \vdots & \vdots & \vdots \\
        \frac{\hat{z}^{2N-1}}{(2N-1)!} & \frac{\hat{z}^{2N-2}}{(2N-2)!} & \cdots &  \frac{\hat{z}^{N}}{N!}
       \end{array} \right)_{2N\times N},
\]
where $\mathbf{H}$ is the $N\times 2N$ matrix given in Eq.~(\ref{defHN}) with $h_j=\theta_j(z_0)$. Utilizing the matrix relation (\ref{PHLH}) as well as Lemma 1, we find that the right side of the above equation is equal to $\mathbf{P}^{-1}\mathbf{L}\mathbf{M}$, where
\[ \label{Mform}
\mathbf{M}=\left( \begin{array}{lll}
         a_{11} +\cdots &0+\cdots & \cdots  \\
         a_{22}\hat{z} +\cdots &a_{22}+\cdots & \cdots \\
        \vdots& \vdots & \vdots  \\
        a_{kk}\frac{\hat{z}^{k-1}}{(k-1)!}  +\cdots  & a_{kk}\frac{\hat{z}^{k-2}}{(k-2)!}  +\cdots  & \cdots  \\
         \beta \frac{\hat{z}^{k+1}}{(k+1)!}+\cdots & \beta \frac{\hat{z}^{k}}{k!}+\cdots & \cdots  \\
         \beta \frac{\hat{z}^{k+3}}{(k+3)!}+\cdots & \beta \frac{\hat{z}^{k+2}}{(k+2)!}+\cdots & \cdots  \\
         \vdots& \vdots & \vdots  \\
         \beta \frac{\hat{z}^{k+1+2(N-k-1)}}{(k+1+2(N-k-1))!} +\cdots & \beta \frac{\hat{z}^{k+2(N-k-1)}}{(k+2(N-k-1))!} +\cdots & \cdots
       \end{array} \right).
\]
In this $\mathbf{M}$ matrix, the terms ``$\cdots$" are terms of higher powers in $\hat{z}$, and each next column of $\mathbf{M}$ is the $\hat{z}$ derivative of its previous column. It is important for us to point out that this form of $\mathbf{M}$ is crucial in our analysis, and Lemma~1 played a critical role in its derivation. Then, using Eqs.~(\ref{AdlerMoserdef}), (\ref{zz0}) and (\ref{thetaH}), we find that
\[
\widehat{\Theta}_{N}(\hat{z})=c_N \det(\mathbf{P})^{-1} \det (\mathbf{M}),
\]
where the fact of $\det(\mathbf{L})=1$ has been utilized since the diagonal elements of the lower triangular matrix $\mathbf{L}$ are all one. The multiplicity of the $\hat{z}=0$ root in $\widehat{\Theta}_{N}(\hat{z})$ is determined by the lowest power term of $\hat{z}$ in $\widehat{\Theta}_{N}(\hat{z})$. This lowest-power term of $\hat{z}$ is obtained by keeping only the first term of each element in the above $\mathbf{M}$ matrix (\ref{Mform}). The determinant of such a reduced $\mathbf{M}$ matrix, that we denote as $\mathbf{M}_0$, can be easily seen as
\begin{eqnarray}
&& \det(\mathbf{M}_0)= \beta^{N-k}\prod_{j=1}^k a_{jj} \det
\left( \begin{array}{cccccc}
         \hat{z} & 1 & 0 & 0 & 0 & \cdots  \\
         \frac{\hat{z}^3}{3!} & \frac{\hat{z}^2}{2!} & \hat{z} & 1  & 0 & \cdots  \\
         \vdots& \vdots & \vdots & \vdots & \vdots & \vdots \\
         \frac{\hat{z}^{2(N-k)-1}}{(2(N-k)-1)!} & \frac{\hat{z}^{2(N-k)-2}}{(2(N-k)-2)!} & \frac{\hat{z}^{2(N-k)-3}}{(2(N-k)-3)!} & \frac{\hat{z}^{2(N-k)-4}}{(2(N-k)-4)!} & \frac{\hat{z}^{2(N-k)-5}}{(2(N-k)-5)!} & \cdots
       \end{array} \right)\\
&& \hspace{1.2cm} = \beta^{N-k}\left(\prod_{j=1}^k a_{jj}\right) c_{N_0}^{-1} \hspace{0.04cm} \hat{z}^{N_0(N_0+1)/2},
\end{eqnarray}
where $N_0=N-k$. Here, the last step was calculated using a technique in \cite{OhtaJY2012}. Thus, the lowest power term of $\hat{z}$ in $\widehat{\Theta}_{N}(\hat{z})$ is proportional to $\hat{z}^{N_0(N_0+1)/2}$, which means that the multiplicity of the $\hat{z}=0$ root in $\widehat{\Theta}_{N}(\hat{z})$, or equivalently, the multiplicity of the $z_0$ root in $\Theta_{N}(z)$, is equal to $N_0(N_0+1)/2$, which is a triangular number. This completes the proof of Theorem 1.  $\Box$

\section{Triangular rogue clusters associated with nonzero multiple roots of Adler--Moser polynomials in the NLS equation} \label{sec:NLS}
The NLS equation
\[ \label{NLS-2020}
\textrm{i} u_{t} +  \frac{1}{2}u_{xx}+ |u|^2 u=0
\]
arises in numerous physical situations such as water waves and optics \cite{Benney,Zakharov,Hasegawa}. Since this equation admits Galilean and scaling invariances, we can set the boundary conditions of its rogue waves as $u(x, t) \to e^{\textrm{i} t}$ as $x, t \to \pm \infty$. Under these boundary conditions, compact expressions of general rogue waves in the NLS equation are given by \cite{Yang2021a}
\[ \label{uNform}
u_N(x,t)=\frac{\sigma_{1}}{\sigma_{0}}e^{\textrm{i}t},
\]
where
\begin{equation} \label{sigma_n}
\sigma_{n}=
\det_{
\begin{subarray}{l}
1\leq i, j \leq N
\end{subarray}
}
\left(
\begin{array}{c}
 \phi_{2i-1,2j-1}^{(n)}
\end{array}
\right),
\end{equation}
\begin{equation} \label{phiijNLS}
\phi_{i,j}^{(n)}=\sum_{\nu=0}^{\min(i,j)} \frac{1}{4^{\nu}} \hspace{0.06cm} S_{i-\nu}(\textbf{\emph{x}}^{+}(n) +\nu \textbf{\emph{s}})  \hspace{0.06cm} S_{j-\nu}(\textbf{\emph{x}}^{-}(n) + \nu \textbf{\emph{s}}),
\end{equation}
vectors $\textbf{\emph{x}}^{\pm}(n)=\left( x_{1}^{\pm}, x_{2}^{\pm},\cdots \right)$ are defined by
\[ \label{defxk}
x_{1}^{\pm}=x \pm \textrm{i} t \pm n, \ \ \ x_{2j}^{\pm} = 0, \quad x_{2j+1}^{+}= \frac{x+2^{2j} (\textrm{i} t)}{(2j+1)!} +a_{2j+1},    \quad x_{2j+1}^{-}= \left(x_{2j+1}^{+}\right)^*,
\]
with $j\ge 1$ and the asterisk * representing complex conjugation, $\textbf{\emph{s}}=(0, s_2, 0, s_4, \cdots)$ are coefficients from the expansion
\[ \label{sexpand}
\sum_{j=1}^{\infty} s_{j}\lambda^{j}=\ln \left[\frac{2}{\lambda}  \tanh \left(\frac{\lambda}{2}\right)\right],
\]
and $a_{3}, a_{5}, \cdots, a_{2N-1}$ are free irreducible complex constants.

When $N=1$, the above solution is $u_1(x,t)=\hat{u}_1(x, t)\hspace{0.04cm} e^{\textrm{i}t}$, where
\begin{equation} \label{Pere}
\hat{u}_1(x, t)=1- \frac{4(1+2\textrm{i}t)}{1+4x^2+4t^2}.
\end{equation}
This is the fundamental rogue wave in the NLS equation that was discovered by Peregrine in \cite{Peregrine1983} and is now called the Peregrine wave in the literature. This wave has a single hump of amplitude 3, flanked by two dips on each side of the $x$ direction. For higher $N$ values and large internal parameters, various rogue patterns would appear.

Patterns of these rogue waves $u_N(x, t)$ under a single large internal parameter $a_{2j+1}$ were studied in our earlier work \cite{Yang2021a}. It was shown that those patterns are predicted by root structures of the Yablonskii--Vorob'ev polynomial hierarchy. If multiple internal parameters in these rogue waves are large and of the single-power form
\[ \label{acond}
a_{2j+1}=\kappa_j \hspace{0.04cm} A^{2j+1}, \quad 1\le j\le N-1,
\]
where $A\gg 1$ is a large positive constant, and $(\kappa_1, \kappa_2, \dots, \kappa_{N-1})$ are $O(1)$ complex constants not being all zero, it was shown in our recent work \cite{YangAD2024} that the corresponding rogue patterns are predicted by the root structure of the Adler--Moser polynomial $\Theta_N(z; \kappa_1, \dots, \kappa_{N-1})$. Specifically, it was shown that if all roots of this Adler--Moser polynomial are simple, then the rogue pattern would comprise fundamental (Peregrine) rogue waves whose locations on the $(x, t)$ plane are proportional to the values of these roots. But if the Adler--Moser polynomial admits multiple roots, the rogue pattern was not resolved in \cite{YangAD2024}. In this case, while each simple root of the Adler--Moser polynomial would still give rise to a Peregrine wave on the $(x, t)$ plane, what wave pattern on the $(x, t)$ plane would be induced by a multiple root is still a key open question. We note that this multiple-root question was considered recently in \cite{Ling2025}, but their multiple large parameters $\{a_{2j+1}\}$ always had additional terms than (\ref{acond}), and their studies were for special coefficients in those large-parameter forms which did not cover our parameter case (\ref{acond}). The focus of this paper is to treat large parameters of the single-power form (\ref{acond}) with arbitrary coefficients $\{\kappa_j\}$.

It turns out that rogue patterns induced by a zero multiple root and a nonzero multiple root are very different. In this section, we treat the case where this multiple root is nonzero. The case of this multiple root being zero will be treated in Sec.~\ref{sec_0root} later.

\subsection{Prediction of a triangular rogue cluster for a nonzero multiple root of the Adler--Moser polynomial}
Now, we consider NLS rogue waves with large internal parameters (\ref{acond}) for general $\{\kappa_j\}$ values. In this case, if the Adler--Moser polynomial $\Theta_N(z; \kappa_1, \dots, \kappa_{N-1})$ admits a nonzero multiple root $z_0$ of multiplicity $N_0(N_0+1)/2$, then we will show that this multiple root would induce a triangular rogue cluster on the $(x, t)$ plane. This cluster comprises $N_0(N_0+1)/2$ Peregrine waves whose $(x, t)$ locations are linearly related to the triangular root structure of the Yablonskii--Vorob'ev polynomial $Q_{N_0}(z)$. Details of these results are presented in the following theorem.

\begin{thm} \label{Theorem2}
For the NLS rogue wave $u_N(x, t)$ with multiple large internal parameters of the single-power form (\ref{acond}), suppose the corresponding Adler--Moser polynomial $\Theta_N(z; \kappa_1, \dots, \kappa_{N-1})$ admits a nonzero multiple root $z_0$ of multplicity $N_0(N_0+1)/2$. Then, a triangular rogue cluster will appear on the $(x, t)$ plane. This rogue cluster comprises $N_0(N_0+1)/2$ Peregrine waves
$\hat{u}_1(x-x_{0}, t-t_{0}) \hspace{0.05cm} e^{{\rm{i}}t}$ forming a triangular shape, where $\hat{u}_1(x, t)$ is given in Eq.~(\ref{Pere}), and positions $(x_{0}, t_{0})$ of these Peregrine waves are given by
\begin{equation}
x_{0}+{\rm{i}}\hspace{0.05cm}t_{0}=z_{0}A + \hat{z}_0 \Omega A^{1/3}, \label{x0t0NLS}
\end{equation}
with $\Omega\equiv \left[-\left(z_0+ 3{\rm i}\Im\left(z_{0} \right)\right)/8\right]^{1/3}$ and $\hat{z}_{0}$ being every one of the $N_0(N_0+1)/2$ simple roots of the Yablonskii--Vorob'ev polynomial $Q_{N_0}(z)$. The error of this Peregrine wave approximation is $O(A^{-1/3})$. Expressed mathematically, when $(x-x_{0})^2+(t-t_{0})^2=O(1)$, we have the following solution asymptotics
\begin{equation} \label{uNpred}
u_{N}(x,t; a_{3}, a_{5}, \cdots, a_{2N-1}) = \hat{u}_1(x-x_{0},t-t_{0})\hspace{0.05cm} e^{\textrm{i}t} + O\left(A^{-1/3}\right).
\end{equation}
\end{thm}

The proof of this theorem will be provided later in this section.

Theorem~\ref{Theorem2} states that the wave pattern induced by a nonzero multiple root of the Adler--Moser polynomial $\Theta_N(z)$ is a triangular rogue cluster. The reason for this triangular shape of the cluster is that the Yablonskii--Vorob'ev polynomial's root structure is triangular (see Fig.~\ref{f:rootYV}). As we can see from Eq.~(\ref{x0t0NLS}), each root $\hat{z}_0$ of the Yablonskii--Vorob'ev polynomial $Q_{N_0}(z)$ gives rise to a Peregrine wave, and positions $(x_0, t_0)$ of these Peregrine waves are given through a linear mapping of $Q_{N_0}(z)$'s root structure (notice that $\Omega$ in Eq.~(\ref{x0t0NLS}) is nonzero when $z_0\ne 0$). Since the root structure of Yablonskii--Vorob'ev polynomials has a triangular shape \cite{Clarkson2003-II,Miller2014,Bertola2016}, their linear mapping is triangular as well. Hence, the rogue cluster is triangular.

\subsection{Numerical verification of the analytical prediction in Theorem~\ref{Theorem2}}

In this subsection, we use two examples to numerically verify the theoretical predictions in Theorem~\ref{Theorem2}.

\begin{figure}[htb]
\begin{center}
\includegraphics[scale=0.40, bb=400 0 350 800]{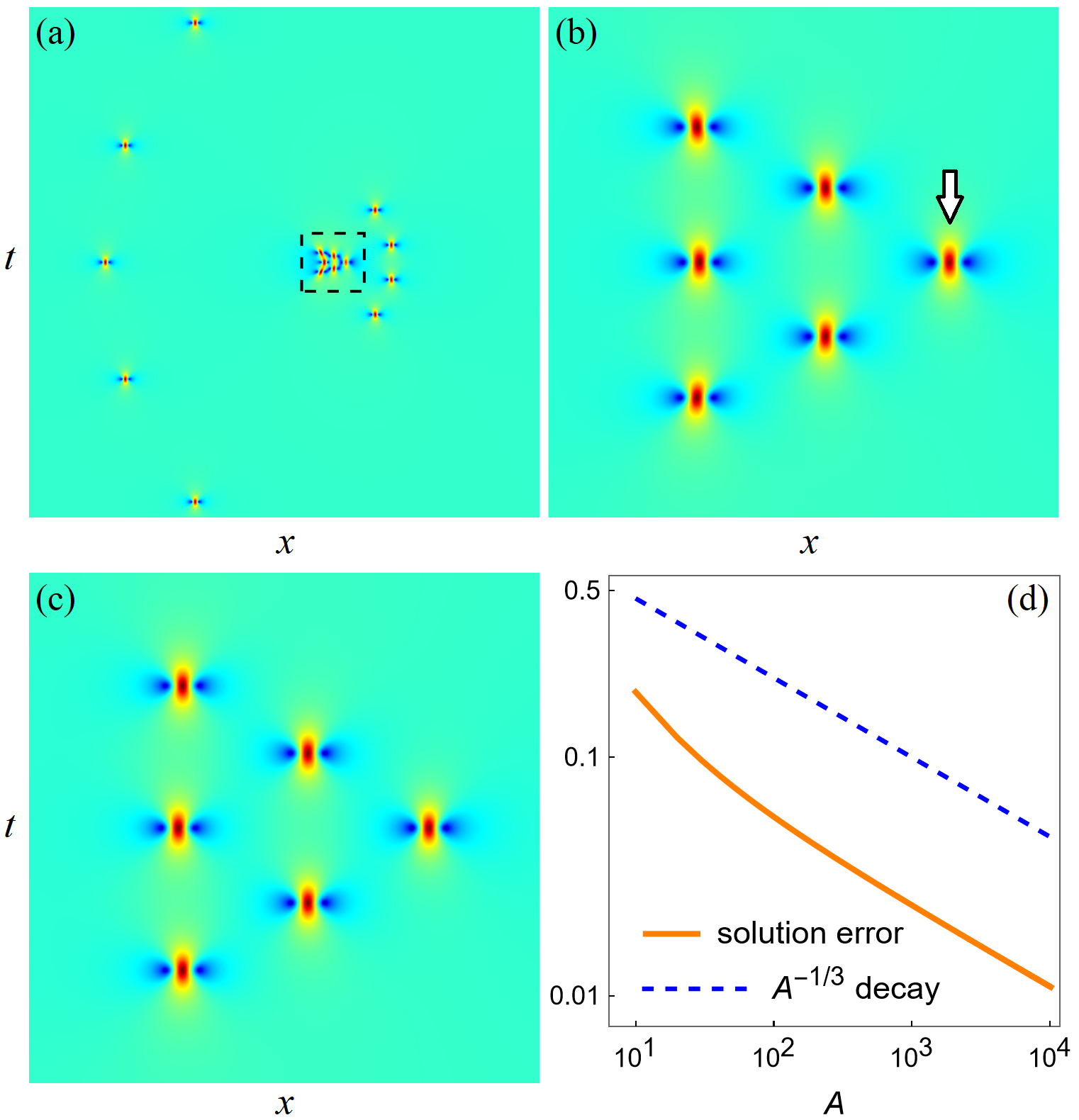}
\caption{(a) A true 5-th order NLS rogue wave $|u_{5}(x, t)|$ for internal parameters (\ref{acond}) with $(\kappa_1, \kappa_2, \kappa_3, \kappa_4)$ given in Eq.~(\ref{parameter1}) and $A=10$. The black dashed box marks the wave cluster of our interest.
(b) Zoom-in of this cluster for a larger $A$ value of $A=200$. (c) Leading-order analytical prediction of this cluster in (b) from Theorem~\ref{Theorem2}. The $(x, t)$ intervals are $ -55 \leq x, t  \leq 55$ for panel (a) and $188 \leq  x \leq 214,\ -13 \leq t \leq 13$ for panels (b) and (c). (d) Error of the leading-order prediction versus $A$ for the Peregrine wave marked by a white arrow in panel (b) (the theoretical decay rate of $A^{-1/3}$ is also plotted for comparison). } \label{f:NLS}
\end{center}
\end{figure}

\textbf{Example 1. } In our first example, we choose $N=5$ and $(\kappa_1, \kappa_2, \kappa_3, \kappa_4)$ as in Eq.~(\ref{parameter1}). When $A=10$, the true rogue wave $u_5(x, t)$ with internal parameters (\ref{acond}) is plotted in Fig.~\ref{f:NLS}(a). It is seen that the wave field contains 9 Peregrine waves forming two opposing arcs, which closely mimic the two arcs of simple roots in the root structure of $\Theta_{5}(z; \kappa_1, \kappa_2, \kappa_{3}, \kappa_{4})$ shown in Fig.~\ref{f:rootAM}(a). This is what we already expected from our earlier work \cite{YangAD2024}. Our current interest is the wave cluster between those two arcs, which is highlighted by a black dashed box in that panel. This wave cluster is associated with the multiple root $z_0=1$ in the root structure of $\Theta_{5}(z; \kappa_1, \kappa_2, \kappa_{3}, \kappa_{4})$ of Fig.~\ref{f:rootAM}(a), which Theorem~\ref{Theorem2} is predicting for parameters (\ref{acond}) with large $A$. This cluster looks triangular with 6 main humps. But these 6 humps are not well separated, thus they are not ready to be compared with Theorem~\ref{Theorem2}'s predictions yet. The reason these 6 humps are not well separated can be understood from Eq.~(\ref{x0t0NLS}) of Theorem~\ref{Theorem2}, which shows that the distances between the predicted Peregrine waves are of $O(A^{1/3})$. Right now, $A=10$, thus these distances are not large, leading to the predicted Peregrine humps staying close together. Theorem~\ref{Theorem2} predicts that better hump separation would be achieved for larger $A$ values. For this reason, we will choose $A=200$ to do the comparison. For this larger $A$ value, the wave cluster corresponding to the multiple root $z_0=1$ is plotted in Fig.~\ref{f:NLS}(b). We see that this cluster is well resolved now, and it comprises 6 well-separated humps forming a triangular pattern, with each hump being an approximate Peregrine wave. In panel (c), we show the leading-order analytical prediction of $|u_5(x, t)|$ in the region of (b) from Theorem~\ref{Theorem2}. Here, the leading-order prediction is a collection of 6 Peregrine waves whose $(x_0, t_0)$ locations are obtained from Eq.~(\ref{x0t0NLS}). We see that this analytical prediction closely resembles the true solution, but differences between them are also clearly visible. To verify the $O(A^{-1/3})$ error decay of our prediction, we show in (d) the error of this prediction versus the $A$ value. Here, the error is measured as the distance between the true and predicted locations of the Peregrine wave marked by a white arrow in panel (b), and the location of the Peregrine wave is numerically determined as the position of its peak amplitude. By comparing this error curve to the theoretical decay rate of $A^{-1/3}$, we see that this error indeed decays as $O(A^{-1/3})$ at large $A$. Thus, Theorem~\ref{Theorem2} is fully confirmed.

\begin{figure}[htb]
\begin{center}
\includegraphics[scale=0.35, bb=500 0 600 400]{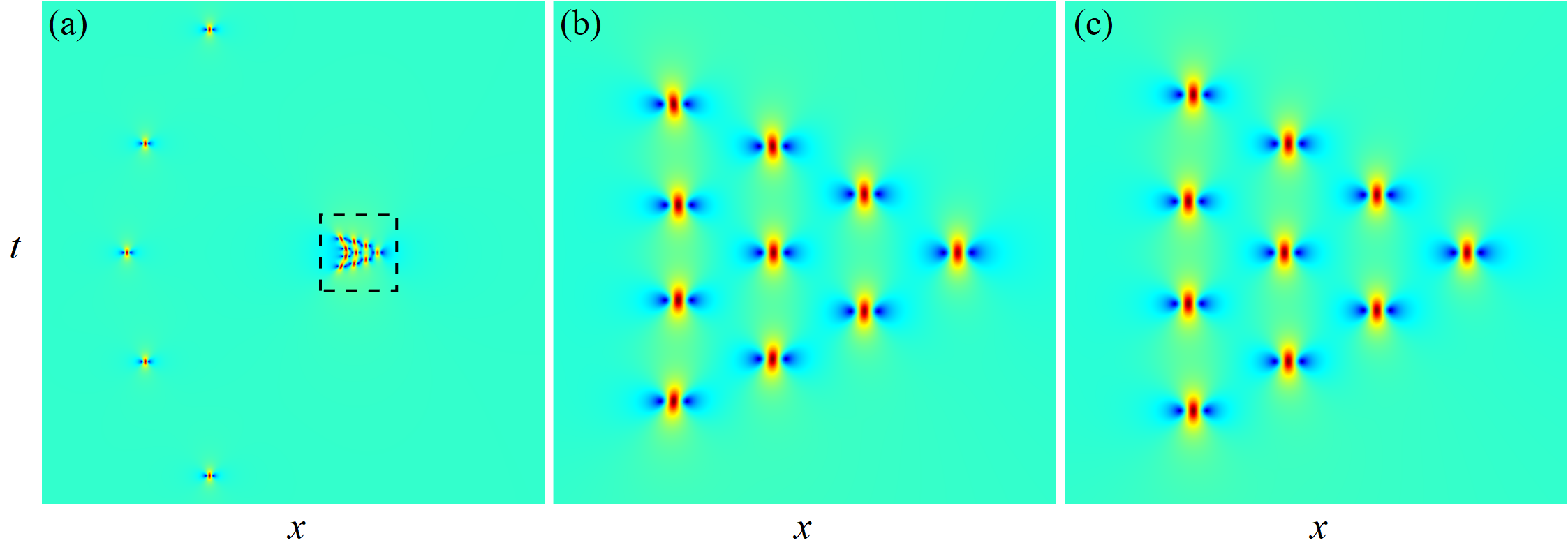}
\caption{(a) A true 5-th order NLS rogue wave $|u_{5}(x, t)|$ for internal parameters (\ref{acond}) with $(\kappa_1, \kappa_2, \kappa_3, \kappa_4)$ given in Eq.~(\ref{parameter2}) and $A=15$. The black dashed box marks the wave cluster of our interest.
(b) Zoom-in of this cluster for a larger $A$ value of $A=200$. (c) Leading-order analytical prediction of this cluster in (b)  from Theorem~\ref{Theorem2}. The $(x, t)$ intervals are $-60 \leq x, t  \leq 60$ for panel (a) and $186 \leq  x \leq 218,\ -16 \leq t \leq 16$ for panels (b) and (c).} \label{f:NLS2}
\end{center}
\end{figure}

\textbf{Example 2. } In our second example, we choose $N=5$ and $(\kappa_1, \kappa_2, \kappa_3, \kappa_4)$ as in Eq.~(\ref{parameter2}). When $A=15$, the true rogue wave $u_5(x, t)$ with internal parameters given in Eq.~(\ref{acond}) is plotted in Fig.~\ref{f:NLS2}(a). It is seen that the left side of the wave field contains 5 Peregrine waves forming an arc, which closely mimics the arc of 5 simple roots in the root structure of $\Theta_{5}(z; \kappa_1, \kappa_2, \kappa_{3}, \kappa_{4})$ shown in Fig.~\ref{f:rootAM}(b), as we would expect from our earlier work \cite{YangAD2024}. Our current interest is the wave cluster on the right side of the wave field, which we have highlighted by a black dashed box in that panel. This cluster is associated with the multiple root $z_0=1$ in the root structure of $\Theta_{5}(z; \kappa_1, \kappa_2, \kappa_{3}, \kappa_{4})$ of Fig.~\ref{f:rootAM}(b), which Theorem~\ref{Theorem2} is predicting. We see that this cluster is triangular with 10 main humps, some of which not well-separated. Thus, to compare this cluster with our predictions, we use a larger value of $A=200$ instead (as we did in Example~1). For this larger $A$ value, the wave cluster corresponding to the multiple root $z_0=1$ is plotted in Fig.~\ref{f:NLS}(b). This cluster comprises 10 well-separated humps forming a triangular pattern, with each hump being an approximate Peregrine wave. In panel (c), we show the leading-order analytical prediction of $|u_5(x, t)|$ in the region of (b) from Theorem~\ref{Theorem2}. We see that this analytical prediction closely resembles the true solution, confirming the predictive power of Theorem~\ref{Theorem2}. We have also verified the $O(A^{-1/3})$ error decay similar to what we did in Example~1, but details will be omitted for brevity.

\subsection{Proof of Theorem~\ref{Theorem2}} \label{sec:NLSproof}
Now, we prove Theorem~\ref{Theorem2} on the asymptotic prediction of a triangular rogue cluster for a nonzero simple root in the Adler--Moser polynomial for the NLS equation.

\textbf{Proof. } We first rewrite the $\sigma_n$ determinant (\ref{sigma_n}) into a $3N\times 3N$ determinant \cite{OhtaJY2012}
\[ \label{sigma3Nby3N}
\sigma_{n}=\left|\begin{array}{cc}
\textbf{O}_{N\times N} & \mathbf{\Phi}_{N\times 2N} \\
-\mathbf{\Psi}_{2N\times N} & \textbf{I}_{2N\times 2N} \end{array}\right|,
\]
where
\[ \label{PhiPsi}
\Phi_{i,j}=2^{-(j-1)} S_{2i-j}\left[\textbf{\emph{x}}^{+} + (j-1) \textbf{\emph{s}}\right], \quad
\Psi_{i,j}=2^{-(i-1)} S_{2j-i}\left[\textbf{\emph{x}}^{-} + (i-1) \textbf{\emph{s}}\right].
\]
To prove Theorem~\ref{Theorem2}, we need to perform asymptotic analysis to this $\sigma_{n}$ determinant for large $A$. For this purpose, we notice that $\textbf{\emph{x}}^{+}=\left(x_1^+, 0, x_{3}^{+}, 0, \cdots\right)$,
where $x_{1}^{+}, x_{3}^{+}, \dots$ are given in Eq.~(\ref{defxk}) which contain internal parameters $a_3, a_5, \dots$.
When these internal parameters are of the form (\ref{acond}) with $A\gg 1$, and $x, t= O(A)$ or smaller, we define $(x_c, t_c)$ by $x_c+{\rm{i}}t_c=z_0A$ and split $\textbf{\emph{x}}^{+}$ as
\begin{eqnarray}
&& \textbf{\emph{x}}^{+}=\textbf{\emph{w}}+\hat{\textbf{\emph{x}}}^{+}, \label{xplussplit} \\
&& \textbf{\emph{w}}\equiv (x_c+{\rm{i}}t_c, 0, a_3, 0, a_5, 0, \dots)=(z_0A, 0, \kappa_1A^3, 0, \kappa_2A^5, 0, \dots),  \\
&& \hat{\textbf{\emph{x}}}^{+}\equiv (\hat{x}_1^{+}, 0, \hat{x}_3^{+}+b_3, 0, \hat{x}_5^{+}+b_5, 0, \dots),  \label{xxkbk} \\
&& \hat{x}_1^{+}\equiv \hat{x}+{\rm{i}}\hat{t}+n, \quad \hat{x}_{2j+1}^{+}\equiv\frac{\hat{x}+2^{2j} (\textrm{i} \hat{t})}{(2j+1)!}, \quad j\ge 1, \\
&& b_{2j+1}\equiv \frac{x_c+2^{2j} (\textrm{i} t_c)}{(2j+1)!}=\frac{z_0+(2^{2j}-1) \Im(z_0)\hspace{0.03cm} \textrm{i} }{(2j+1)!}A, \quad j\ge 1,  \label{b2j1} \\
&& \hat{x}\equiv x-x_c, \quad \hat{t}\equiv t-t_c,
\end{eqnarray}
where $\Re$ and $\Im$ represent the real and imaginary parts of a complex number. From this $\textbf{\emph{x}}^{+}$ splitting and the definition of Schur polynomials, we see that
\[ \label{Sjsplit}
S_{j}(\textbf{\emph{x}}^{+} +\nu \textbf{\emph{s}}) = \sum_{i=0}^j S_{j-i}(\textbf{\emph{w}})S_{i}(\hat{\textbf{\emph{x}}}^{+} +\nu \textbf{\emph{s}}),
\]
where $\nu$ is any integer. In addition, from the definition of functions $\theta_{j}(z)$ in Eq.~(\ref{AMthetak}), we see that
\[ \label{Sjw}
S_j(\textbf{\emph{w}})=A^j \theta_j(z_0).
\]
Let us denote $\theta_j(z_0)\equiv h_j$. Then, using the above two relations, we can rewrite the $\mathbf{\Phi}$ matrix as
\begin{eqnarray}
&& \mathbf{\Phi}=\mathbf{F}\mathbf{G},  \label{PhiFG} \\
&& \mathbf{F}=\mbox{Mat}_{1\le i\le N,\hspace{0.06cm} 1\le j\le 2N}\left(A^{2i-j}h_{2i-j}\right)= \mathbf{D}_1\hspace{0.02cm} \mathbf{H} \hspace{0.04cm} \mathbf{D}_2,  \\
&& \mathbf{G}=\mbox{Mat}_{1\le i, j\le 2N}\left(
2^{-(j-1)} S_{i-j}\left[\hat{\textbf{\emph{x}}}^{+} + (j-1) \textbf{\emph{s}}\right]\right), \\
&& \mathbf{D}_1=\mbox{diag}(A, A^3, \dots, A^{2N-1}), \\
&& \mathbf{D}_2=\mbox{diag}(1, A^{-1}, \dots, A^{1-2N}).
\end{eqnarray}
Here, $\mathbf{H}$ is the matrix given in Eq.~(\ref{defHN}) with $h_j=\theta_j(z_0)$, which is the same $\mathbf{H}$ matrix as in the proof of Theorem~1. From Eq.~(\ref{PHLH}) and Lemma~1, we get $\mathbf{H}=\mathbf{P}^{-1}\mathbf{L}\widehat{\mathbf{H}}$, where $\mathbf{P}$ is a permutation matrix, $\mathbf{L}$ is an upper triangular matrix with nonzero diagonal elements, and $\widehat{\mathbf{H}}$ is a row echelon form given in Eq.~(\ref{defHhatN}). Thus,
\[
\mathbf{\Phi}=\mathbf{D}_1\hspace{0.02cm} \mathbf{P}^{-1}\mathbf{L}\widehat{\mathbf{H}} \hspace{0.04cm} \mathbf{D}_2 \hspace{0.01cm}\mathbf{G}.
\]
The key step of this proof is to utilize the special row echelon form $\widehat{\mathbf{H}}$ in Eq.~(\ref{defHhatN}) of Lemma~1. Doing so, we find that we can rewrite the above $\mathbf{\Phi}$ matrix as
\[ \label{PhiM}
\mathbf{\Phi}=\mathbf{D}_1\hspace{0.02cm} \mathbf{P}^{-1}\mathbf{L}\mathbf{D}_3\mathbf{M},
\]
where $\mathbf{D}_3=\mbox{diag}(1, A^{-1}, \dots, A^{1-k}, A^{-1-k}, A^{-3-k}, \dots, A^{1-2(N-k)-k})$,
\[ \label{defMNLS}
\mathbf{M}= \left(
\begin{array}{lll}  a_{1,1}S_{0}+a_{1,2}S_{1}A^{-1}+\cdots & \frac{1}{2}(0+a_{1,2}S_{0}A^{-1}+\cdots) & \cdots
  \\ a_{2,2}S_{1}+a_{2,3}S_2A^{-1}+\cdots & \frac{1}{2}(a_{2,2}S_{0}+a_{2,3}S_{1}A^{-1}+\cdots ) & \cdots
\\ \vdots & \vdots & \vdots \\ a_{k,k}S_{k-1}+a_{k,k+1}S_{k}A^{-1}+\cdots &
\frac{1}{2}(a_{k,k}S_{k-2}+a_{k,k+1}S_{k-1}A^{-1}+\cdots) & \cdots \\
\beta S_{k+1}+b_{1,3}S_{k+2}A^{-1}+\cdots &
\frac{1}{2}(\beta S_{k}+b_{1,3}S_{k+1}A^{-1}+\cdots) & \cdots
\\ \beta S_{k+3}+b_{2,5}S_{k+4}A^{-1}+\cdots &
\frac{1}{2}(\beta S_{k+2}+b_{2,5}S_{k+3}A^{-1}+\cdots) & \cdots \\
\vdots & \vdots & \vdots \\
\beta S_{k-1+2N_0}+b_{N_0,2N_0+1}S_{k+2N_0}A^{-1}+\cdots &
\frac{1}{2}(\beta S_{k-2+2N_0}+b_{N_0,2N_0+1}S_{k-1+2N_0}A^{-1}+\cdots) & \cdots
\end{array}
\right)_{N\times 2N},
\]
$k$ is the number of the first few columns of $\mathbf{H}$ that are linearly independent, $N_0=N-k$,
$a_{i,j}$ and $b_{i,j}$ are elements in the $\mathbf{A}$ and $\mathbf{B}$ matrices of Lemma~1, and
$S_{j}$ is equal to $S_{j}\left(\hat{\textbf{\emph{x}}}^{+}\right)$ in the first column, equal to $S_{j}\left(\hat{\textbf{\emph{x}}}^{+}+\textbf{\emph{s}}\right)$ in the second column, and so on.
Notice that the determinant of the $N\times N$ matrix comprising the first $N$ columns of $\mathbf{H}$ is just $\Theta_{N}(z_0)$, the fact of $z_0$ being a root of $\Theta_{N}(z)$ means that these first $N$ columns of $\mathbf{H}$ are linearly dependent. Thus, $k<N$ and $N_0\ge 1$. This $\mathbf{M}$ matrix above is similar to a matrix of the same name in the proof of Theorem~1. Since the $\mathbf{H}$ matrix here (with $h_j=\theta_j(z_0)$) is the same as that in the proof of Theorem~1, we see that $N_0(N_0+1)/2$ is the multiplicity of the $z_0$ root in the Adler--Moser polynomial $\Theta_N(z; \kappa_1, \dots, \kappa_{N-1})$.

Matrices $\mathbf{D}_1$, $\mathbf{P}^{-1}$, $\mathbf{L}$ and $\mathbf{D}_3$ in Eq.~(\ref{PhiM}) are all $N\times N$ nonsingular constant matrices that are independent of the index $n$ of $\sigma_n$. Because of that, they can all be factored out of the $\sigma_n$ determinant (\ref{sigma3Nby3N}) and cancel out from $\sigma_1/\sigma_0$ in the rogue wave formula (\ref{uNform}). This means that in the $\sigma_n$ determinant (\ref{sigma3Nby3N}), $\mathbf{\Phi}$ can be replaced by $\mathbf{M}$. Similarly, $\mathbf{\Psi}$ in that $\sigma_n$ determinant can be replaced by a counterpart matrix $\widetilde{\mathbf{M}}$ of $\mathbf{M}$. Thus, the asymptotics of $\sigma_n$ can be obtained from analyzing the asymptotics of $\mathbf{M}$ and its counterpart $\widetilde{\mathbf{M}}$.

To proceed further, we will first use a heuristic approach to derive the leading order approximation of $u_N(x, t)$. Afterwards, we will use a more rigorous analysis to justify that leading order approximation and derive its error estimates.

Our heuristic approach is as follows. As we will quickly see, rogue patterns induced by the multiple root $z_0$ of the Adler--Moser polynomial $\Theta_N(z)$ appear in the $(\hat{x}, \hat{t})=O(A^{1/3})$ region of the $(x, t)$ plane. In this region, since $(\hat{x}_{1}^{+}, \hat{x}_{3}^{+}, \dots)$ are all $O(A^{1/3})$ and $(b_3, b_5, \dots)$ all $O(A)$, the expression for $\hat{\textbf{\emph{x}}}^{+}$ in Eq.~(\ref{xxkbk}) indicates that
\[ \label{Sjasym}
S_{j}(\hat{\textbf{\emph{x}}}^{+} +\nu \textbf{\emph{s}})=O(A^{j/3})
\]
for any fixed integer $\nu$. Thus, we see from Eq.~(\ref{defMNLS}) that at large $A$, all terms involving $A^{-1}$ or its powers in the $\mathbf{M}$ matrix are subdominant, and
\[ \label{MM0}
\mathbf{M}\sim \mathbf{M}_0,   \quad A\gg 1,
\]
where $\mathbf{M}_0$ is the matrix of $\mathbf{M}$ with all terms involving $A^{-1}$ and its powers neglected, i.e.,
\[ \label{defM0NLS}
\mathbf{M}_0=\left(\begin{array}{cc} \mathbf{M}_a & 0    \\  \mathbf{M}_c & \mathbf{M}_b
\end{array} \right),
\]
\[
\mathbf{M}_a= \left(
\begin{array}{cccc}  a_{1,1}S_{0} & 0 & 0 & \cdots
  \\ a_{2,2}S_{1} & \frac{1}{2} a_{2,2}S_{0} & 0 & \cdots
\\ \vdots & \vdots & \ddots & \vdots \\ a_{k,k}S_{k-1}&
\frac{1}{2}a_{k,k}S_{k-2} & \cdots & \frac{1}{2^{k-1}}a_{k,k}S_{0}
\end{array}
\right)_{k\times k},
\]
\[
\mathbf{M}_b=\beta\hspace{0.04cm} \mbox{Mat}_{1\le i \le N-k, 1\le j \le 2N-k}\left(2^{-(j-1)-k} S_{2i-j}\left[\hat{\textbf{\emph{x}}}^{+} + (j-1+k) \textbf{\emph{s}}\right]\right),
\]
and $\mathbf{M}_c$ is a matrix we do not write out since it is not needed. Note that in the above lower-triangular matrix $\mathbf{M}_a$, $S_{j}$ is equal to $S_{j}\left(\hat{\textbf{\emph{x}}}^{+}\right)$ in the first column, equal to $S_{j}\left(\hat{\textbf{\emph{x}}}^{+}+\textbf{\emph{s}}\right)$ in the second column, and so on.

Using the above results and their counterparts for the $\mathbf{\Psi}$ component, and recalling that $a_{1,1}, a_{2,2}, \dots, a_{k,k}$ and $\beta$ are all nonzero and $S_0=1$, we see that the $\sigma_n$ determinant (\ref{sigma3Nby3N}) with its $\mathbf{\Phi}$ replaced by $\mathbf{M}$ and its $\mathbf{\Psi}$ replaced by $\mathbf{M}$'s counterpart $\widetilde{\mathbf{M}}$ is asymptotically equal to
\[ \label{sigmasigmahat}
\sigma_n\sim \alpha_0 \hspace{0.04cm} \hat{\sigma}_n,       \quad A\gg 1,
\]
where $\alpha_0$ is a certain nonzero constant,
\[ \label{sigmahat}
\hat{\sigma}_n=\left|\begin{array}{cc}
\textbf{O}_{N_0\times N_0} & \widehat{\mathbf{\Phi}}_{N_0\times 2N_0} \\
-\widehat{\mathbf{\Psi}}_{2N_0\times N_0} & \textbf{I}_{2N_0\times 2N_0} \end{array}\right|,
\]
and
\[ \label{PhiPsihat}
\widehat{\Phi}_{i,j}=2^{-(j-1)} S_{2i-j}\left[\hat{\textbf{\emph{x}}}^{+} + (j-1+k) \textbf{\emph{s}}\right], \quad
\widehat{\Psi}_{i,j}=2^{-(i-1)} S_{2j-i}\left[\hat{\textbf{\emph{x}}}^{-} + (i-1+k) \textbf{\emph{s}}\right].
\]
During this calculation, an overall factor of $2^{-k}$ has been scaled out from the $\mathbf{M}_b$ matrix and its counterpart for the $\mathbf{\Psi}$ component. Using techniques of Ref.~\cite{Yang2021a}, we can remove the $k$ term in the above equation (\ref{PhiPsihat}) without affecting the $\hat{\sigma}_n$ determinant. Then, the resulting $\hat{\sigma}_n$ simply corresponds to the $N_0$-th order rogue wave $u_{N_0}(\hat{x}, \hat{t})$ with internal parameters $(b_3, b_5, \dots, b_{2N_0-1})$. Thus, we have
\[ \label{uNuNhat}
u_N(x, t; a_3, a_5, \dots, a_{2N-1})\sim u_{N_0}(\hat{x}, \hat{t}; b_3, b_5, \dots, b_{2N_0-1})e^{{\rm{i}}(t-\hat{t})}, \quad A\gg 1.
\]
The phase term $e^{{\rm{i}}(t-\hat{t})}$ here is induced by our notation in Eq.~(\ref{uNform}), which implies $u_N(x, t)$ has phase $e^{{\rm{i}}t}$ while $u_{N_0}(\hat{x}, \hat{t})$ has phase $e^{{\rm{i}}\hat{t}}$. From Eq.~(\ref{b2j1}), we see that internal $b_{2j+1}$ parameters in this $N_0$-th order rogue wave $u_{N_0}(\hat{x}, \hat{t})$ are nonzero and $O(A)$. The asymptotics of this $u_{N_0}(\hat{x}, \hat{t})$ has been studied in Sec.~6 of Ref.~\cite{Yang2021a}. Since the current internal parameters satisfy the condition of $b_{2j+1}\le O(b_3^{(2j-1)/3})$ for every $j\ge 2$, results in Sec.~6 of Ref.~\cite{Yang2021a} indicate that at large $A$, the solution $u_{N_0}(\hat{x}, \hat{t}; b_3, b_5, \dots, b_{2N_0-1})$ would split into $N_0(N_0+1)/2$ Peregrine waves $\hat{u}_1(\hat{x}-\hat{x}_{0}, \hat{t}-\hat{t}_{0}) \hspace{0.05cm} e^{{\rm{i}}\hat{t}}$, where $\hat{u}_1(x, t)$ is given in Eq.~(\ref{Pere}), and positions $(\hat{x}_{0}, \hat{t}_{0})$ of these Peregrine waves are given by $\hat{x}_{0}+{\rm{i}}\hspace{0.05cm}\hat{t}_{0}=\hat{z}_0\left(-3b_3/4\right)^{1/3}$, with $\hat{z}_{0}$ being every one of the $N_0(N_0+1)/2$ simple roots of the Yablonskii--Vorob'ev polynomial $Q_{N_0}(z)$. Since $b_3=[z_0+3\Im(z_0) \hspace{0.03cm}\textrm{i}]A/6$,
we then get
\begin{equation} \label{x0t0hat}
\hat{x}_{0}+{\rm{i}}\hspace{0.05cm}\hat{t}_{0}=\hat{z}_0 \hspace{0.03cm} \Omega \hspace{0.03cm} A^{1/3},
\end{equation}
where $\Omega$ is as defined in Theorem~\ref{Theorem2} and is nonzero. Recalling $\hat{x}=x-x_c$ and $\hat{t}=t-t_c$ with $x_c+{\rm{i}}t_c=z_0A$, we see that $\hat{u}_1(\hat{x}-\hat{x}_{0}, \hat{t}-\hat{t}_{0})=\hat{u}_1(x-x_{0}, t-t_{0})$, where $(x_0, t_0)$ are as given in Eq.~(\ref{x0t0NLS}). Then, Eq.~(\ref{uNuNhat}) means that
\[ \label{uNasym9}
u_N(x, t; a_3, a_5, \dots, a_{2N-1})\sim \hat{u}_1(x-x_{0},t-t_{0})\hspace{0.05cm} e^{\textrm{i}t}, \quad A\gg 1
\]
when $(x, t)$ are in the $O(1)$ neighborhood of $(x_0, t_0)$.

The above derivation is heuristic for the following reason. The leading-order term $\hat{\sigma}_n$ in Eq.~(\ref{sigmasigmahat}) turns out to nearly vanish around locations (\ref{x0t0hat}) where Peregrine waves are predicted. This fact can be seen from Ref.~\cite{Yang2021a} or from the later text of this subsection. Because of that, it is crucial for us to show that the error terms which are neglected in the leading-order asymptotics (\ref{sigmasigmahat}) do not surpass or match that leading-order contribution in those regions. Since we did not estimate those errors and their relative contributions, the above calculation was heuristic and not rigorous.

Next, we more carefully justify the above asymptotics (\ref{uNasym9}) and derive its error estimates. In this process, we will not rely on our earlier results in Ref.~\cite{Yang2021a}, but will do all necessary calculations directly so that our treatment here is self-contained.

First, we split the $\mathbf{M}$ matrix in Eq.~(\ref{defMNLS}) as
\[ \label{Msplit}
\mathbf{M}= \mathbf{M}_0+\mathbf{M}_1,
\]
where $\mathbf{M}_0$ is as given in Eq.~(\ref{defM0NLS}). We also do a similar splitting for the counterpart matrix $\widetilde{\mathbf{M}}$ of the $\mathbf{\Psi}$ counterpart. In the $(\hat{x}, \hat{t})=O(A^{1/3})$ region, due to the asymptotics (\ref{Sjasym}), we see that when $(\mathbf{M}_{0})_{ij}\ne 0$, the matrix element $(\mathbf{M}_{1})_{ij}$ of $\mathbf{M}_1$ is $O(A^{-2/3})$ less than the corresponding matrix element $(\mathbf{M}_{0})_{ij}$ of $\mathbf{M}_0$; and when $(\mathbf{M}_{0})_{ij}=0$, $(\mathbf{M}_{1})_{ij}$ is order of $A^{-1}$ or its higher power. Similar results hold for the counterpart $\widetilde{\mathbf{M}}$ in the $\mathbf{\Psi}$ component.

Then, we examine the matrix $\mathbf{M}_0$. This matrix comprises elements $S_j\left(\hat{\textbf{\emph{x}}}^{+} + \nu\textbf{\emph{s}}\right)$. When $(\hat{x}, \hat{t})=O(A^{1/3})$, it is easy to see that
\[\label{Skasym2}
S_{j}(\hat{\textbf{\emph{x}}}^{+} +\nu \textbf{\emph{s}}) = S_{j}\left(\hat{x}_1^+, 0, b_3, 0, 0,  \cdots\right)\left[1+O(A^{-2/3})\right]
\]
for any fixed integer $\nu$.
The polynomial $S_{j}\left(\hat{x}_1^+, 0, b_3, 0, 0,  \cdots\right)$ is related to $p_{j}(z)$ in Eq.~(\ref{defpk}) as
\begin{equation} \label{Skorder2}
S_{j}\left(\hat{x}_1^+, 0, b_3, 0, 0,  \cdots\right)=\Omega^j A^{j/3}p_{j}(\hat{z}),
\end{equation}
where $\Omega$ is as defined in Theorem~\ref{Theorem2}, and $\hat{z}=\Omega^{-1} A^{-1/3}(\hat{x}+\textrm{i}\hat{t}+n)$. Inserting (\ref{Skorder2}) into (\ref{Skasym2}), we get
\[\label{Skasym3}
S_{j}(\hat{\textbf{\emph{x}}}^{+} +\nu \textbf{\emph{s}}) = \Omega^j A^{j/3}p_{j}(\hat{z})\left[1+O\left(A^{-2/3}\right)\right].
\]

Now, we use the above results (\ref{Msplit}), (\ref{Skasym3}) and their $\mathbf{\Psi}$-counterparts to calculate $\sigma_n$ in Eq. (\ref{sigma3Nby3N}), with its $\mathbf{\Phi}$ replaced by $\mathbf{M}$ and its $\mathbf{\Psi}$ replaced by $\mathbf{M}$'s counterpart $\widetilde{\mathbf{M}}$. To proceed, we first use determinant identities and the Laplace expansion to rewrite that $\sigma_n$ as
\[ \label{sigmanLap}
\sigma_{n}=\sum_{1\leq\mu_{1} < \mu_{2} < \cdots < \mu_{N}\leq 2N}
\det_{1 \leq i, j\leq N} \left(\mathbf{M}_{i,\mu_j}\right)\times \det_{1 \leq i, j\leq N} \left(\widetilde{\mathbf{M}}_{\mu_j,i}\right).
\]
It is easy to see that the dominant contributions to this $\sigma_n$ come from two index choices, one being $\mu=(1, 2, \cdots, N)$, and the other being $\mu=(1, 2, \cdots, N-1, N+1)$, and the rest of the contributions are of relative order $A^{-1/3}$.

With the first index choice, in view of Eqs. (\ref{Msplit}), (\ref{Skasym3}) and size discussions of $\mathbf{M}_{1}$'s elements above, the $\mathbf{M}_{i,\mu_j}$ determinant in Eq.~(\ref{sigmanLap}) can be found as
\[ \label{Phinnxtnew}
\alpha_1 \hspace{0.06cm} A^{\frac{N_0(N_0+1)}{6}} \left[Q_{N_0}(\hat{z})+O\left(A^{-2/3}\right)\right],
\]
where $\alpha_1=2^{-N(N-1)/2}\Omega^{N_0(N_0+1)/2}\beta^{N_0}c_{N_0}^{-1}\prod_{i=1}^k a_{ii}$. Here, the leading-order contribution to this determinant comes from approximating $\mathbf{M}$ by $\mathbf{M}_0$ and approximating $S_{j}(\hat{\textbf{\emph{x}}}^{+} +\nu \textbf{\emph{s}})$ in $\mathbf{M}_0$ by its leading-order term in Eq.~(\ref{Skasym3}), and the $O(A^{-2/3})$ error term in (\ref{Phinnxtnew}) comes from the $\mathbf{M}_1$ component of $\mathbf{M}$ as well as the $O(A^{-2/3})$ error term in (\ref{Skasym3}). In view of the definitions of $(\hat{x}_{0}, \hat{t}_{0})$ in Eq.~(\ref{x0t0hat}), we can rewrite $\hat{z}$ as
\[ \label{hatzexp}
\hat{z}=\hat{z}_0+\Omega^{-1} A^{-1/3}\left[(\hat{x}-\hat{x}_0)+\textrm{i}(\hat{t}-\hat{t}_0)+n\right].
\]
Then, expanding $Q_{N_0}(\hat{z})$ around $\hat{z}=\hat{z}_0$, and recalling $\hat{z}_0$ is
a simple root of the Yablonskii--Vorob'ev polynomial $Q_{N_0}(z)$, i.e., $Q_{N_0}(\hat{z}_0)=0$ and $Q'_{N_0}(\hat{z}_0)\ne 0$, we get
\[
Q_{N_0}(\hat{z})=\Omega^{-1}A^{-1/3}Q'_{N_0}(\hat{z}_0)\left[(\hat{x}-\hat{x}_{0})+\textrm{i}(\hat{t}-\hat{t}_{0})+n\right]
\left[1+O\left(A^{-1/3}\right)\right].
\]
Inserting this equation into (\ref{Phinnxtnew}), the $\mathbf{M}_{i,\mu_j}$ determinant in Eq.~(\ref{sigmanLap}) then becomes
\[
\alpha_1 \Omega^{-1}Q'_{N_0}(\hat{z}_0) \left[(\hat{x}-\hat{x}_{0})+\textrm{i}(\hat{t}-\hat{t}_{0})+n\right] A^{\frac{N_0(N_0+1)-2}{6}}\left[1+O\left(A^{-1/3}\right)\right].
\]
Similarly, the $\widetilde{\mathbf{M}}_{\mu_j,i}$ determinant in Eq.~(\ref{sigmanLap}) can be found as
\[
\left(\alpha_1 \Omega^{-1}Q'_{N_0}(\hat{z}_0)\right)^*\left[(\hat{x}-\hat{x}_{0})-\textrm{i}(\hat{t}-\hat{t}_{0})-n\right] \hspace{0.06cm} A^{\frac{N_0(N_0+1)-2}{6}}\left[1+O\left(A^{-1/3}\right)\right].
\]

With the second index choice of $\mu=(1, 2, \cdots, N-1, N+1)$, the leading-order contribution to the $\mathbf{M}_{i,\mu_j}$ determinant in Eq.~(\ref{sigmanLap}) can be obtained by neglecting the $\mathbf{M}_1$ component of $\mathbf{M}$ and approximating $S_{j}(\hat{\textbf{\emph{x}}}^{+} +\nu \textbf{\emph{s}})$ in $\mathbf{M}_0$ by its leading-order term in Eq.~(\ref{Skasym3}), and the relative error of this approximation is $O(A^{-2/3})$. Thus, this $\mathbf{M}_{i,\mu_j}$ determinant is found as
\[
\frac{1}{2}\alpha_1 c_{N_0} \Omega^{-1}
\det_{1 \leq i \leq N_0} \left[p_{2i-1}(\hat{z}),  p_{2i-2}(\hat{z}),
\cdots, p_{2i-(N_0-1)}(\hat{z}),  p_{2i-(N_0+1)}(\hat{z})\right]A^{\frac{N_0(N_0+1)-2}{6}} \left[1+O\left(A^{-2/3}\right)\right].
\]
Since $p_{2i-(N_0+1)}(\hat{z})=p'_{2i-N}(\hat{z})$, this determinant is then equal to
\[ \label{Phinnxtnew2}
\frac{1}{2}\alpha_1 \Omega^{-1} Q'_{N_0}(\hat{z}) A^{\frac{N_0(N_0+1)-2}{6}} \left[1+O\left(A^{-2/3}\right)\right].
\]
Utilizing Eq.~(\ref{hatzexp}), this expression can be approximated as
\[
\frac{1}{2}\alpha_1 \hspace{0.06cm} \Omega^{-1} Q'_{N_0}(\hat{z}_0) A^{\frac{N_0(N_0+1)-2}{6}} \left[1+O\left(A^{-1/3}\right)\right].
\]
Similarly, the $\widetilde{\mathbf{M}}_{\mu_j,i}$ determinant in Eq.~(\ref{sigmanLap}) can be found as
\[
\frac{1}{2}\left(\alpha_1 \hspace{0.06cm} \Omega^{-1} Q'_{N_0}(\hat{z}_0) \right)^* A^{\frac{N_0(N_0+1)-2}{6}} \left[1+O\left(A^{-1/3}\right)\right].
\]

Summarizing the above contributions, we find that the determinant $\sigma_n$ in Eq.~(\ref{sigmanLap}) is calculated as
\[ \label{sigmanxt5}
\sigma_{n} =\left|\alpha_1\Omega^{-1}\right|^2 \hspace{0.06cm} \left|Q'_{N_0}(\hat{z}_0)\right|^2 A^{\frac{N_0(N_0+1)-2}{3}}
\left[ \left(\hat{x}-\hat{x}_{0}\right)^2+\left(\hat{t}-\hat{t}_{0}\right)^2 - (2 \textrm{i}) n \left(\hat{t}-\hat{t}_{0}\right)-n^2+\frac{1}{4} \right]\left[1+O\left(A^{-1/3}\right)\right].
\]
Then, inserting the above asymptotics into Eq.~(\ref{uNform}), we find that when $(\hat{x}, \hat{t})$ is in the $O(1)$ neighborhood of $\left(\hat{x}_{0}, \hat{t}_{0}\right)$, i.e., when $(x, t)$ is in the $O(1)$ neighborhood of $(x_{0}, t_{0})$ where $(x_0, t_0)$ are given in Eq.~(\ref{x0t0NLS}) of Theorem~\ref{Theorem2},
\[
u_N(x, t; a_3, a_5, \dots, a_{2N-1}) = \frac{\sigma_{1}}{\sigma_{0}}e^{\textrm{i}t} =e^{\textrm{i}t}
\left(1- \frac{4[1+2\textrm{i}(\hat{t}-\hat{t}_{0})]}{1+4(\hat{x}-\hat{x}_{0})^2+4(\hat{t}-\hat{t}_{0})^2}\right) + O\left(A^{-1/3}\right),
\]
which is a Peregrine wave $\hat{u}_1(x-x_{0},t-t_{0})\hspace{0.05cm} e^{\textrm{i}t}$, and the error of this Peregrine approximation is $O\left(A^{-1/3}\right)$. Theorem~\ref{Theorem2} is then proved.  $\Box$

\section{Triangular rogue clusters in the GDNLS equations}

The normalized GDNLS equations are \cite{Kundu1984,Clarkson1987,Satsuma_GDNLS_soliton,YangDNLS2020}
\[ \label{GDNLS}
\textrm{i}u_t+\frac{1}{2}u_{xx}+\textrm{i}\gamma |u|^2u_x+\textrm{i}(\gamma-1)u^2u_x^*+\frac{1}{2}(\gamma-1)(\gamma-2)|u|^4u=0,
\]
where $\gamma$ is a real constant. These equations become the Kaup-Newell equation when $\gamma=2$ \cite{Kaup_Newell}, the Chen-Lee-Liu equation when $\gamma=1$ \cite{CCL}, and the Gerdjikov-Ivanov equation when $\gamma=0$ \cite{GI}. These GDNLS equations and their special versions govern a number of physical processes such as the propagation of circularly polarized nonlinear Alfv\'en waves in plasmas \cite{KN_Alfven1,KN_Alfven2}, short-pulse propagation in a frequency-doubling crystal \cite{Wise2007}, and propagation of ultrashort pulses in a single-mode optical fiber \cite{HasegawaKodama1995,Agrawal_book}.

Rogue waves in these equations satisfy the following normalized boundary conditions \cite{YangDNLS2020}
\begin{eqnarray}
&& u(x,t)\rightarrow  e^{{\rm i}(1-\gamma-\alpha) x-\frac{1}{2}{\rm{i}}[\alpha^2+2(\gamma-2)\alpha+1-\gamma]t},  \quad x, t \to \pm \infty, \label{BoundaryCond1}
\end{eqnarray}
where $\alpha>0$ is a free wave number parameter. Under these conditions, compact expressions of $N$-th order rogue waves in the GDNLS equations (\ref{GDNLS}) are given by \cite{Yang2021b}.
\[ \label{uNGDNLS}
u_N(x,t)= e^{{\rm i}(1-\gamma-\alpha) x
-\frac{1}{2}{\rm{i}}[\alpha^2+2(\gamma-2)\alpha+1-\gamma]t}
\hspace{0.05cm}\frac{(f_N^*)^{\gamma-1}g_N}{f_N^\gamma},
\]
where
\begin{equation} \label{fNgNGDNLS}
f_N(x,t)=\sigma_{0,0}, \quad g_N(x,t)=\sigma_{-1,1},
\end{equation}
\begin{equation} \label{sigma_nD}
\sigma_{n,k}=
\det_{\begin{subarray}{l}
1\leq i, j \leq N
\end{subarray}}
\left(\begin{array}{c}
\phi_{2i-1,2j-1}^{(n,k)}
\end{array}\right),
\end{equation}
\[ \label{matrixmnij}
\phi_{i,j}^{(n,k)}=\sum_{\nu=0}^{\min(i,j)} \frac{1}{4^{\nu}} \hspace{0.06cm} S_{i-\nu}(\textbf{\emph{x}}^{+}(n,k) +\nu \textbf{\emph{s}})  \hspace{0.06cm} S_{j-\nu}(\textbf{\emph{x}}^{-}(n,k) + \nu \textbf{\emph{s}}),
\]
the vectors $\textbf{\emph{x}}^{\pm}(n,k)=\left( x_{1}^{\pm}, x_{2}^{\pm},\cdots \right)$ are defined by
\begin{eqnarray}
&&x_{1}^{+}=\hspace{0.2cm}   k+ \left(n+\frac{1}{2}\right)\left(h_{1}+\frac{1}{2}\right)+\sqrt{\alpha}\hspace{0.04cm} x+ \left[\sqrt{\alpha}(\alpha-1)+ {\rm i}\alpha\right]t,  \label{defx1GDNLS} \\
&&x_{1}^{-}=-k- \left(n+\frac{1}{2}\right)\left(h_{1}^*+\frac{1}{2}\right)+\sqrt{\alpha}\hspace{0.04cm} x+ \left[\sqrt{\alpha}(\alpha-1)-{\rm i}\alpha\right]t, \\
&&x_{2j+1}^{+}=\hspace{0.25cm} (n+\frac{1}{2}) \hspace{0.04cm} h_{2j+1}+ \frac{1}{(2j+1)!}\left\{\sqrt{\alpha}\hspace{0.04cm} x+\left[\sqrt{\alpha}(\alpha-1)+2^{2j} {\rm i} \alpha\right]t\right\} +a_{2j+1},  \quad j\ge 1,  \\
&&x_{2j+1}^{-}= -(n+\frac{1}{2}) \hspace{0.04cm} h_{2j+1}^*+ \frac{1}{(2j+1)!}\left\{\sqrt{\alpha}\hspace{0.04cm} x+\left[ \sqrt{\alpha}(\alpha-1)-2^{2j} {\rm i} \alpha\right]t\right\} + a_{2j+1}^*, \quad j\ge 1, \\
&& x_{2j}^{\pm}=0, \quad j\ge 1,
\end{eqnarray}
$\emph{\textbf{s}}=(s_{1}, s_{2}, \cdots)$ is defined in Eq.~(\ref{sexpand}),
$h_{r}(\alpha)$ are coefficients from the expansion
\[
\sum_{j=1}^{\infty} h_{j}\lambda^{j}=\ln \left(\frac{{\rm i}e^{\lambda/2}+\sqrt{\alpha} e^{-\lambda/2}}{{\rm i}+\sqrt{\alpha}}\right),
\]
and $a_{3}, a_{5}, \cdots, a_{2N-1}$ are free irreducible complex constants.

The fundamental GDNLS rogue wave is obtained when we take $N=1$ in the above general solution. This fundamental rogue wave is
\[ \label{u1GDNLS}
u_1(x, t)=\hat{u}_1(x, t)\hspace{0.04cm} e^{{\rm i}(1-\gamma-\alpha) x
-\frac{1}{2}{\rm{i}}[\alpha^2+2(\gamma-2)\alpha+1-\gamma]t},
\]
where
\[ \label{u1hatGDNLS}
\hat{u}_1(x, t)=\frac{(f_1^*)^{\gamma-1}g_1}{f_1^\gamma},
\]
the functions $f_1(x,t)$ and $g_1(x,t)$ are given from Eq.~(\ref{fNgNGDNLS}) as
\begin{eqnarray}
f_1(x,t)=&& \hspace{-0.5cm} \left[\frac{1}{2}\left(h_{1}+\frac{1}{2}\right)+\sqrt{\alpha}\hspace{0.04cm} x+ \left[\sqrt{\alpha}(\alpha-1)+ {\rm i}\alpha\right]t\right] \nonumber \\
&& \hspace{-0.5cm} \times  \left[-\frac{1}{2}\left(h_{1}^*+\frac{1}{2}\right)+\sqrt{\alpha}\hspace{0.04cm} x+ \left[\sqrt{\alpha}(\alpha-1)-{\rm i}\alpha\right]t \right]+\frac{1}{4}, \label{f1} \\
g_1(x,t)=&& \hspace{-0.5cm} \left[1-\frac{1}{2}\left(h_{1}+\frac{1}{2}\right)+\sqrt{\alpha}\hspace{0.04cm} x+ \left[\sqrt{\alpha}(\alpha-1)+ {\rm i}\alpha\right]t\right] \nonumber \\
&& \hspace{-0.5cm} \times  \left[-1+\frac{1}{2}\left(h_{1}^*+\frac{1}{2}\right)+\sqrt{\alpha}\hspace{0.04cm} x+ \left[\sqrt{\alpha}(\alpha-1)-{\rm i}\alpha\right]t \right]+\frac{1}{4}, \label{g1}
\end{eqnarray}
and $h_1=[{\rm i}-\sqrt{\alpha}]/[2({\rm i}+\sqrt{\alpha})]$. This wave has a single hump of amplitude 3, flanked by two dips on its sides, and its intensity profile is slanted on the $(x, t)$ plane.

Patterns of these GDNLS rogue waves $u_N(x, t)$ under a single large internal parameter $a_{2j+1}$ were studied in our earlier work \cite{Yang2021b,Yangbook2024}. It was shown that those patterns are predicted by root structures of the Yablonskii--Vorob'ev polynomial hierarchy. If multiple internal parameters in these rogue waves are large and of the single-power form (\ref{acond}), it was shown in \cite{Yangbook2024} that the corresponding rogue patterns are predicted by the root structure of the Adler--Moser polynomial $\Theta_N(z; \kappa_1, \dots, \kappa_{N-1})$. If all roots of this Adler--Moser polynomial are simple, then the rogue pattern would comprise fundamental rogue waves whose locations on the $(x, t)$ plane are a certain linear transformation to this polynomial's root structure. When the Adler--Moser polynomial admits multiple roots, while each simple root of the polynomial would still give rise to a fundamental rogue wave on the $(x, t)$ plane, the wave pattern induced by a multiple root was not addressed in \cite{Yangbook2024}. This question will be answered in this paper. Similar to the NLS case, rogue patterns induced by a zero multiple root and a nonzero multiple root are very different. In this section, we treat the nonzero multiple-root case.  The zero multiple-root case will be treated in Sec.~\ref{sec_0root} later.

\subsection{Prediction of a triangular rogue cluster and its proof for the GDNLS equations}
In this section, we consider GDNLS rogue waves with large internal parameters (\ref{acond}) when the corresponding Adler--Moser polynomial $\Theta_N(z; \kappa_1, \dots, \kappa_{N-1})$ admits a nonzero multiple root. If this root has multiplicity $N_0(N_0+1)/2$, then we will show that this root would induce a triangular rogue cluster on the $(x, t)$ plane. This cluster comprises $N_0(N_0+1)/2$ fundamental rogue waves whose $(x, t)$ locations are linearly related to the triangular root structure of the Yablonskii--Vorob'ev polynomial $Q_{N_0}(z)$. Details of our results are presented in the following theorem.

\begin{thm} \label{Theorem3}
For the GDNLS rogue wave $u_N(x, t)$ with multiple large internal parameters of the single-power form (\ref{acond}), suppose the corresponding Adler--Moser polynomial $\Theta_N(z; \kappa_1, \dots, \kappa_{N-1})$ admits a nonzero multiple root $z_0$ of multplicity $N_0(N_0+1)/2$. Then, a triangular rogue cluster will appear on the $(x, t)$ plane. This rogue cluster comprises $N_0(N_0+1)/2$ fundamental rogue waves $\hat{u}_1(x-x_{0}, t-t_{0})\hspace{0.06cm} e^{{\rm i}(1-\gamma-\alpha) x-\frac{1}{2}{\rm{i}}\left[\alpha^2+2(\gamma-2)\alpha+1-\gamma\right]t}$ forming a triangular shape, where $\hat{u}_1(x, t)$ is given in Eq.~(\ref{u1hatGDNLS}), and positions $(x_{0}, t_{0})$ of these fundamental rogue waves are given by
\[ \label{x0t0GDNLS}
\sqrt{\alpha}\hspace{0.04cm} x_0+ \left[\sqrt{\alpha}(\alpha-1)+ {\rm i}\alpha\right]t_0=z_{0}A + \hat{z}_0 \Omega A^{1/3},
\]
with $\Omega\equiv \left[-\left(z_0+ 3{\rm i}\Im\left[z_{0} \right]\right)/8\right]^{1/3}$ and $\hat{z}_{0}$ being every one of the $N_0(N_0+1)/2$ simple roots of the Yablonskii--Vorob'ev polynomial $Q_{N_0}(z)$. The error of this fundamental rogue wave approximation is $O(A^{-1/3})$. Expressed mathematically, when $(x-x_{0})^2+(t-t_{0})^2=O(1)$, we have the following solution asymptotics
\begin{equation}
u_{N}(x,t; a_{3}, a_{5}, \cdots, a_{2N-1}) = \hat{u}_1(x-x_{0},t-t_{0})\hspace{0.05cm} e^{{\rm i}(1-\gamma-\alpha) x-\frac{1}{2}{\rm{i}}[\alpha^2+2(\gamma-2)\alpha+1-\gamma]t} + O\left(A^{-1/3}\right).
\end{equation}
\end{thm}

This theorem says that the wave pattern induced by a nonzero multiple root of the Adler--Moser polynomial $\Theta_N(z)$ is a triangular rogue cluster, similar to the NLS case. The reason for this triangular shape of the cluster is easy to see from Eq.~(\ref{x0t0GDNLS}). This equation shows that positions $(x_0, t_0)$ of fundamental rogue waves in this cluster are given through a linear mapping of the root structure $\hat{z}_0$ of the Yablonskii--Vorob'ev polynomial $Q_{N_0}(z)$. Indeed, this linear mapping can be worked out more explicitly from Eq.~(\ref{x0t0GDNLS}) as
\[ \label{mapGDNLS}
\left[\begin{array}{c} x_0 \\ t_0 \end{array}\right]=
\left[\begin{array}{c} x_c \\ t_c \end{array}\right]+
\textbf{B}\left[\begin{array}{c} \Re(\hat{z}_0) \\ \Im(\hat{z}_0) \end{array}\right],
\]
where
\[ \label{xctcGDNLS}
\left[\begin{array}{c} x_c \\ t_c \end{array}\right]=\textbf{B}_0 \left[\begin{array}{c} \Re(z_0) \\ \Im(z_0) \end{array}\right],
\]
and
\[ \label{B0GDNLS}
\textbf{B}_0=A\left[\begin{array}{cc} \frac{1}{\sqrt{\alpha}} & -\frac{\alpha-1}{\alpha}  \\ 0 & \frac{1}{\alpha} \end{array}\right], \quad
\textbf{B}=A^{1/3}\left[\begin{array}{cc} \frac{1}{\sqrt{\alpha}}\Re(\Omega)-\frac{\alpha-1}{\alpha}\Im(\Omega) & -\frac{1}{\sqrt{\alpha}}\Im(\Omega)-\frac{\alpha-1}{\alpha}\Re(\Omega)  \\ \frac{1}{\alpha}\Im(\Omega) & \frac{1}{\alpha}\Re(\Omega) \end{array}\right].
\]
In this linear map (\ref{mapGDNLS}), the first term is a constant shift, and $\textbf{B}$ is a constant matrix. Since the root structure of all Yablonskii--Vorob'ev polynomials has a triangular shape \cite{Clarkson2003-II,Miller2014,Bertola2016}, the rogue cluster of fundamental rogue waves obtained through this linear mapping is triangular as well.

\textbf{Proof of Theorem~\ref{Theorem3}. } The proof of this theorem is similar to that for the NLS equation in Sec.~\ref{sec:NLSproof} and thus will only be sketched below.

We start by rewriting the $\sigma_n$ determinant (\ref{sigma_nD}) into the $3N\times 3N$ determinant (\ref{sigma3Nby3N}), where
$\mathbf{\Phi}$ and $\mathbf{\Psi}$ are given by Eq.~(\ref{PhiPsi}), except that the $\textbf{\emph{x}}^{\pm}$ vectors are now different. Due to the expression of $x_1^+$ in Eq.~(\ref{defx1GDNLS}), we define $(x_c, t_c)$ by
\[ \label{xctcz0AGDNLS}
\sqrt{\alpha}\hspace{0.04cm} x_c+ \left[\sqrt{\alpha}(\alpha-1)+ {\rm i}\alpha\right]t_c=z_0A.
\]
Explicit expressions of $(x_c, t_c)$ can be easily worked out and they are as given in Eq.~(\ref{xctcGDNLS}).

Next, we split $\textbf{\emph{x}}^{+}$ as
\begin{eqnarray}
&& \textbf{\emph{x}}^{+}=\textbf{\emph{w}}+\hat{\textbf{\emph{x}}}^{+}, \\
&& \textbf{\emph{w}}\equiv (\sqrt{\alpha}\hspace{0.04cm} x_c+[\sqrt{\alpha}(\alpha-1)+ {\rm i}\alpha]t_c, 0, a_3, 0, a_5, 0, \dots)=(z_0A, 0, \kappa_1A^3, 0, \kappa_2A^5, 0, \dots),  \\
&& \hat{\textbf{\emph{x}}}^{+}\equiv (\hat{x}_1^{+}, 0, \hat{x}_3^{+}+b_3, 0, \hat{x}_5^{+}+b_5, 0, \dots),  \\
&& \hat{x}_1^{+}\equiv k+ \left(n+\frac{1}{2}\right)\left(h_{1}+\frac{1}{2}\right)+\sqrt{\alpha}\hspace{0.04cm} \hat{x}+ \left[\sqrt{\alpha}(\alpha-1)+ {\rm i}\alpha\right]\hat{t}, \\
&& \hat{x}_{2j+1}^{+}=(n+\frac{1}{2}) \hspace{0.04cm} h_{2j+1}+ \frac{1}{(2j+1)!}\left\{\sqrt{\alpha}\hspace{0.04cm} \hat{x}+\left[\sqrt{\alpha}(\alpha-1)+2^{2j} {\rm i} \alpha\right]\hat{t}\right\}, \quad j\ge 1, \\
&& b_{2j+1}\equiv \frac{1}{(2j+1)!}\left\{\sqrt{\alpha}\hspace{0.04cm} x_c+\left[\sqrt{\alpha}(\alpha-1)+2^{2j} {\rm i} \alpha\right]t_c\right\} =\frac{z_0+(2^{2j}-1) \Im(z_0)\hspace{0.03cm} \textrm{i} }{(2j+1)!}A, \quad j\ge 1,  \label{b2j1GDNLS}\\
&& \hat{x}\equiv x-x_c, \quad \hat{t}\equiv t-t_c.
\end{eqnarray}
Notice that these $b_{2j+1}$ values have the same final expressions as those in Eq.~(\ref{b2j1}) of the NLS case. The rest of the calculations is almost identical to that in the proof of Theorem~\ref{Theorem2} for the NLS equation, the reason being that the rogue solution's structure (\ref{sigma_nD})-(\ref{matrixmnij}) of the GDNLS equations is identical to (\ref{sigma_n})-(\ref{phiijNLS}) of the NLS equation. Based on the heuristic arguments over there, we similarly find that
\[\label{uNuNhatGD}
u_N(x, t; a_3, a_5, \dots, a_{2N-1})\sim u_{N_0}(\hat{x}, \hat{t}; b_3, b_5, \dots, b_{2N_0-1})
e^{{\rm i}(1-\gamma-\alpha) (x-\hat{x})-\frac{1}{2}{\rm{i}}[\alpha^2+2(\gamma-2)\alpha+1-\gamma](t-\hat{t})}, \quad A\gg 1.
\]
Since internal $b_{2j+1}$ parameters in this $N_0$-th order rogue wave $u_{N_0}(\hat{x}, \hat{t})$ are nonzero and $O(A)$, the asymptotics of this rogue wave $u_{N_0}(\hat{x}, \hat{t})$ can be obtained from Ref.~\cite{Yang2021b} and Sec.~6 of Ref.~\cite{Yang2021a} with very little modification, and we find that at large $A$, this $u_{N_0}(\hat{x}, \hat{t})$ would split into $N_0(N_0+1)/2$ fundamental rogue waves $\hat{u}_1(\hat{x}-\hat{x}_{0}, \hat{t}-\hat{t}_{0})\hspace{0.06cm} e^{{\rm i}(1-\gamma-\alpha) \hat{x}-\frac{1}{2}{\rm{i}}[\alpha^2+2(\gamma-2)\alpha+1-\gamma]\hat{t}}$, where $\hat{u}_1(x, t)$ is given in Eq.~(\ref{u1hatGDNLS}), and its  positions $(\hat{x}_{0}, \hat{t}_{0})$ are given by
\[
\sqrt{\alpha}\hspace{0.04cm} \hat{x}_{0}+ \left[\sqrt{\alpha}(\alpha-1)+ {\rm i}\alpha\right]\hat{t}_{0}
=\hat{z}_0\left(-3b_3/4\right)^{1/3}=\hat{z}_0\Omega A^{1/3},
\]
with $\hat{z}_{0}$ being every one of the $N_0(N_0+1)/2$ simple roots of the Yablonskii--Vorob'ev polynomial $Q_{N_0}(z)$ and
$\Omega$ as defined in Theorem~\ref{Theorem3}. Recalling $\hat{x}=x-x_c$ and $\hat{t}=t-t_c$, we see that $\hat{u}_1(\hat{x}-\hat{x}_{0}, \hat{t}-\hat{t}_{0})=\hat{u}_1(x-x_{0}, t-t_{0})$, where $(x_0, t_0)$ are as given in Eq.~(\ref{x0t0GDNLS}), and Eq.~(\ref{uNuNhatGD}) then becomes
\[
u_N(x, t; a_3, a_5, \dots, a_{2N-1})\sim \hat{u}_1(x-x_{0}, t-t_{0})\hspace{0.06cm} e^{{\rm i}(1-\gamma-\alpha) x-\frac{1}{2}{\rm{i}}[\alpha^2+2(\gamma-2)\alpha+1-\gamma]t}, \quad A\gg 1
\]
when $(x, t)$ are in the $O(1)$ neighborhood of $(x_0, t_0)$. Error estimates to the above asymptotics can be obtained in the same way as in the proof of Theorem~\ref{Theorem2}, and we can see that this error is $O(A^{-1/3})$. Theorem~\ref{Theorem3} is then proved. $\Box$

\subsection{Numerical verification of analytical predictions in Theorem~\ref{Theorem3}}

Next, we use an example to numerically verify the theoretical predictions in Theorem~\ref{Theorem3} for the GDNLS equations.

{\textbf{Example 3. } In our example, we choose $\gamma=2$, $\alpha=16/9$, $N=5$, and $(\kappa_1, \kappa_2, \kappa_3, \kappa_4)$ as in Eq.~(\ref{parameter1}). When $A=8$, the true rogue wave $u_5(x, t)$ with internal parameters given in Eq.~(\ref{acond}) is plotted in Fig.~\ref{f:GDNLS}(a). It is seen that the wave field contains two opposing arcs comprising 5 and 4 fundamental GDNLS rogue waves each. These fundamental rogue waves are induced by simple roots in the root structure of $\Theta_{5}(z; \kappa_1, \kappa_2, \kappa_{3}, \kappa_{4})$ shown in Fig.~\ref{f:rootAM}(a), as has been explained in our earlier work \cite{Yangbook2024}. Our current interest is the wave cluster between those two arcs, which is induced by the multiple root $z_0=1$ in the root structure of $\Theta_{5}(z; \kappa_1, \kappa_2, \kappa_{3}, \kappa_{4})$ in Fig.~\ref{f:rootAM}(a). This cluster comprises 6 humps forming a triangle, but some of those 6 humps are not well separated. As done before, we will choose a larger $A$ value of $A=150$ to do the comparison between the true solution and Theorem~\ref{Theorem3}'s predictions. For this larger $A$ value, the wave cluster corresponding to the multiple root $z_0=1$ is plotted in Fig.~\ref{f:GDNLS}(b). We see that this cluster comprises 6 well-separated humps forming a triangular pattern, with each hump being an approximate fundamental rogue wave. In panel (c), we show the leading-order analytical prediction of $|u_5(x, t)|$ in the region of (b) from Theorem~\ref{Theorem3}. Here, the leading-order prediction is a collection of 6 fundamental rogue waves whose $(x_0, t_0)$ locations are obtained from Eq.~(\ref{x0t0GDNLS}). We see that this analytical prediction closely resembles the true solution. To verify the $O(A^{-1/3})$ error decay of our prediction, we show in (d) the error of this prediction versus the $A$ value. Here, the error is measured as the distance between the true and predicted locations of the fundamental rogue wave marked by a white arrow in panel (b), and the location of the fundamental rogue wave is numerically determined as the position of its peak amplitude. By comparing this error curve to the theoretical decay rate of $A^{-1/3}$, we see that this error indeed decays as $O(A^{-1/3})$ at large $A$. Thus, Theorem~\ref{Theorem3} is fully confirmed.

\begin{figure}[htb]
\begin{center}
\includegraphics[scale=0.40, bb=400 0 350 800]{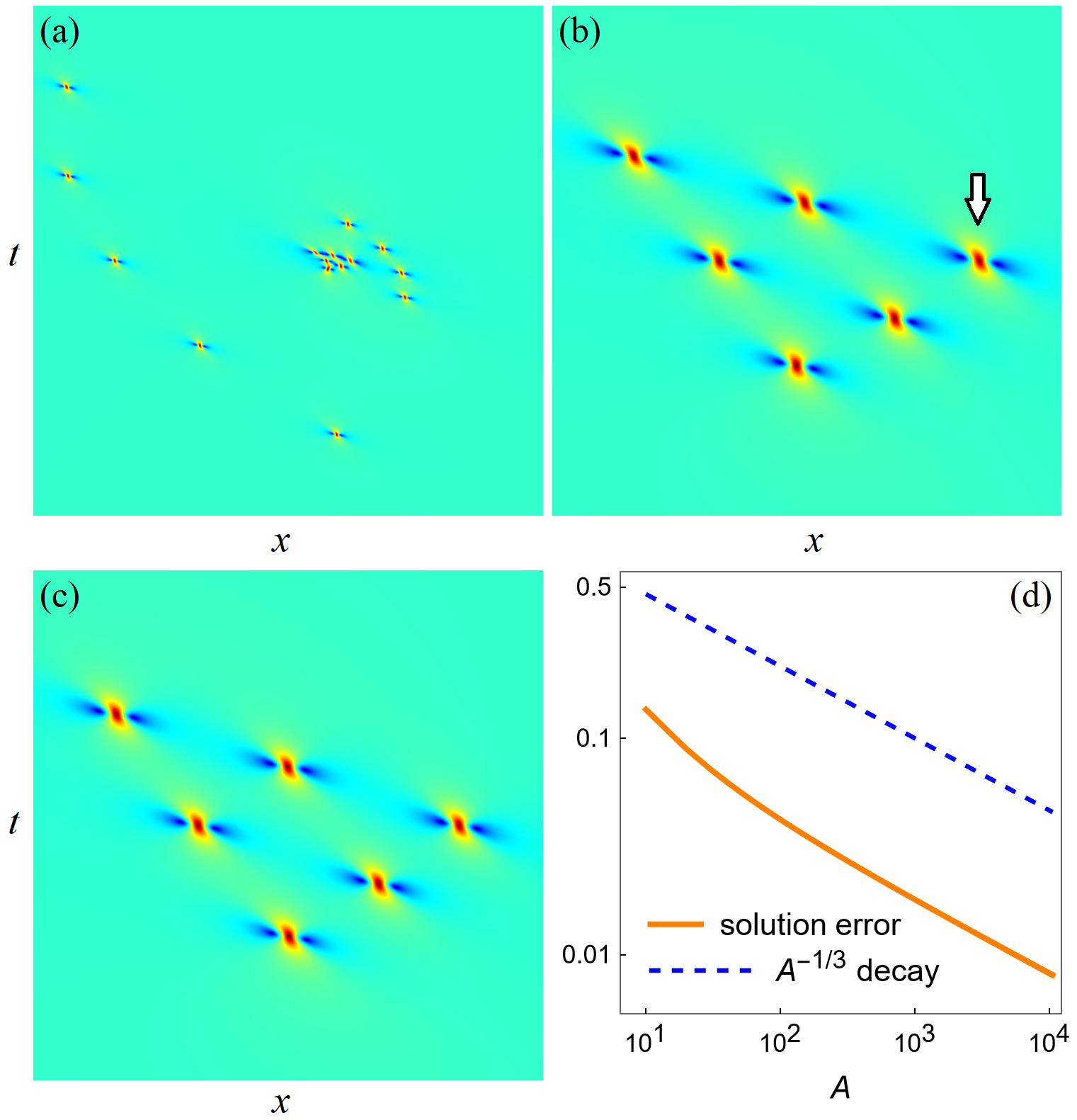}
\caption{(a) A true 5-th order GDNLS rogue wave $|u_{5}(x, t)|$ with $\gamma=2$ and $\alpha=16/9$ for internal parameters (\ref{acond}) with $(\kappa_1, \kappa_2, \kappa_3, \kappa_4)$ given in Eq.~(\ref{parameter1}) and $A=8$. (b) Zoom-in of the wave cluster induced by the multiple root $z_0=1$ of $\Theta_{5}(z; \kappa_1, \kappa_2, \kappa_{3}, \kappa_{4})$ for a larger $A$ value of $A=150$. (c) Leading-order analytical prediction of this cluster in (b) from Theorem~\ref{Theorem3}. The $(x, t)$ intervals are $ -34 \leq x, t  \leq 34$ for panel (a) and $104 \leq  x \leq 121,\ -8.5 \leq t \leq 8.5$ for panels (b) and (c). (d) Error of the leading-order prediction versus $A$ for the fundamental rogue wave marked by a white arrow in panel (b) (the theoretical decay rate of $A^{-1/3}$ is also plotted for comparison). } \label{f:GDNLS}
\end{center}
\end{figure}

\section{Rogue patterns associated with a zero multiple root in the Adler--Moser polynomial} \label{sec_0root}
In the past two sections, we determined rogue patterns induced by a nonzero multiple root in the Adler--Moser polynomial for the NLS and GDNLS equations. We showed that in both cases, a triangular rogue cluster would appear. If the multiple root of the Adler--Moser polynomial is zero, the situation would be totally different. In this case, we will show that instead of a triangular rogue cluster, a lower-order rogue wave would appear in the $O(1)$ neighborhood of the spatial-temporal origin. Details of these results for the NLS and GDNLS equations are presented in the following theorem.

\begin{thm} \label{Theorem4}
For the NLS rogue wave $u_N(x, t)$ in Eq.~(\ref{uNform}) and the GDNLS rogue wave $u_N(x, t)$ in Eq.~(\ref{uNGDNLS}) with multiple large internal parameters of the single-power form (\ref{acond}), suppose the corresponding Adler--Moser polynomial $\Theta_N(z; \kappa_1, \dots, \kappa_{N-1})$ admits a zero multiple root of multplicity $N_0(N_0+1)/2$. Then, a lower $N_0$-th order rogue wave with all internal parameters zero would appear in the $O(1)$ neighborhood of the origin $(x,t)=(0,0)$, and the error of this lower-order rogue wave approximation is $O(A^{-1})$. Expressed mathematically, when $x^2+t^2=O(1)$, we have the following rogue-wave asymptotics for the NLS and GDNLS equations,
\begin{equation}  \label{uNz0zero}
u_{N}(x,t; a_{3}, a_{5}, \cdots, a_{2N-1}) = u_{N_0}(x,t; 0, 0, \cdots, 0)\hspace{0.05cm}  + O\left(A^{-1}\right).
\end{equation}
\end{thm}

\textbf{Proof. } These proofs for the NLS and GDNLS equations are almost identical. Thus, we will just prove it for the NLS equation below.

Suppose $\Theta_N(z; \kappa_1, \dots, \kappa_{N-1})$ admits a zero multiple root of multplicity $N_0(N_0+1)/2$, and $x^2+t^2=O(1)$. We first rewrite the $\sigma_n$ determinant (\ref{sigma_n}) into a $3N\times 3N$ determinant (\ref{sigma3Nby3N}). Then, we split $\textbf{\emph{x}}^{+}$ as
\begin{eqnarray}
&& \textbf{\emph{x}}^{+}=\textbf{\emph{w}}+\hat{\textbf{\emph{x}}}^{+}, \\
&& \textbf{\emph{w}}\equiv (0, 0, a_3, 0, a_5, 0, \dots)=(0, 0, \kappa_1A^3, 0, \kappa_2A^5, 0, \dots),  \\
&& \hat{\textbf{\emph{x}}}^{+}\equiv (\hat{x}_1^{+}, 0, \hat{x}_3^{+}, 0, \hat{x}_5^{+}, 0, \dots),  \label{xxkbk2} \\
&& \hat{x}_1^{+}\equiv x+{\rm{i}}t+n, \quad \hat{x}_{2j+1}^{+}\equiv\frac{x+2^{2j} (\textrm{i} t)}{(2j+1)!}, \quad j\ge 1.
\end{eqnarray}
This splitting is a special case of the earlier (\ref{xplussplit}) with $z_0=0$. Then, using the formulae (\ref{Sjsplit})-(\ref{Sjw}) as well as the special row echelon form in Eq.~(\ref{defHhatN}) of Lemma~1, we can rewrite the $\mathbf{\Phi}$ matrix in (\ref{sigma3Nby3N}) as (\ref{PhiM}), where $\mathbf{M}$ is given in Eq.~(\ref{defMNLS}) except that the $\hat{\textbf{\emph{x}}}^{+}$ vector in $S_{j}\left(\hat{\textbf{\emph{x}}}^{+}+\nu \textbf{\emph{s}}\right)$ there should be updated to (\ref{xxkbk2}) now. The fact of zero being a root of $\Theta_{N}(z)$ guarantees that $k<N$ and $N_0\ge 1$ in the $\mathbf{M}$ matrix (\ref{defMNLS}). When $x^2+t^2=O(1)$, $S_{j}(\hat{\textbf{\emph{x}}}^{+} +\nu \textbf{\emph{s}})=O(1)$. Thus, we still have the asymptotics (\ref{MM0}), i.e., $\mathbf{M}\sim \mathbf{M}_0$, where $\mathbf{M}_0$ is given in Eq.~(\ref{defM0NLS}). From this asymptotics, we still get Eq.~(\ref{sigmasigmahat}), i.e., $\sigma_n\sim \alpha_0 \hspace{0.04cm} \hat{\sigma}_n$, where $\hat{\sigma}_n$ is given in Eq.~(\ref{sigmahat}). Using techniques of Ref.~\cite{Yang2021a}, we can also remove the $k$ term in Eq.~(\ref{PhiPsihat}). Then, we see from Eq.~(\ref{xxkbk2}) that the resulting $\hat{\sigma}_n$ now corresponds to the $N_0$-th order rogue wave $u_{N_0}(x, t)$ with all-zero internal parameters, i.e.,
\[ \label{uNuNhat2}
u_N(x, t; a_3, a_5, \dots, a_{2N-1})\sim u_{N_0}(x, t; 0, 0, \dots, 0), \quad A\gg 1.
\]
Unlike the $z_0\ne 0$ case in the proof of Theorem~\ref{Theorem2}, the leading-order term $\hat{\sigma}_n$ of $\sigma_n$
here does not vanish in the $x^2+t^2=O(1)$ region. Thus, the above argument is no-longer heuristic but is reliable. Regarding the order of error in the approximation (\ref{uNuNhat2}), since $S_{j}(\hat{\textbf{\emph{x}}}^{+} +\nu \textbf{\emph{s}})=O(1)$ in the $\mathbf{M}$ matrix (\ref{defMNLS}), we see that the approximation of $\mathbf{M}$ by $\mathbf{M}_0$ has relative error of $O(A^{-1})$. This translates to an error of $O(A^{-1})$ in the approximation (\ref{uNuNhat2}) as well, hence Eq.~(\ref{uNz0zero}) holds. This completes the proof of Theorem~\ref{Theorem4}. $\Box$

Next, we use a NLS example to confirm Theorem~\ref{Theorem4}. In this example, we take $N=5$ and $(\kappa_1,\kappa_2, \kappa_3, \kappa_4)=(1, 1, 1, 4/3)$ as in Eq.~(\ref{parameter3}). In this case, zero is a triple root of the Adler--Moser polynomial $\Theta_5(z; \kappa_1,\kappa_2, \kappa_3, \kappa_{4})$, see Fig.~\ref{f:rootAM}(c). When we take large internal parameters as (\ref{acond}) in the NLS equation with $A=6$, the true rogue wave is plotted in Fig.~\ref{f:NLSz0zero}(a). The $(x, t)=O(1)$ region that is associated with the zero root of the Adler--Moser polynomial is zoomed in and shown in panel (b). In (c), the analytical prediction for this region from Theorem~\ref{Theorem4}
is displayed. This analytical prediction is a second-order rogue wave with zero internal parameters. As one can see, the predicted solution closely resembles the true solution in panel (b). In panel (d), the error of our approximation versus $A$ is plotted. Here, the error is measured as the absolute difference between the true solution and the predicted solution at the spatial-temporal location of $(x, t)=(0.5, 0.5)$. It can be seen that this error decays in proportion to $A^{-1}$, which matches our prediction in Theorem~\ref{Theorem4}.

\begin{figure}[htb]
\begin{center}
\includegraphics[scale=0.40, bb=400 0 350 820]{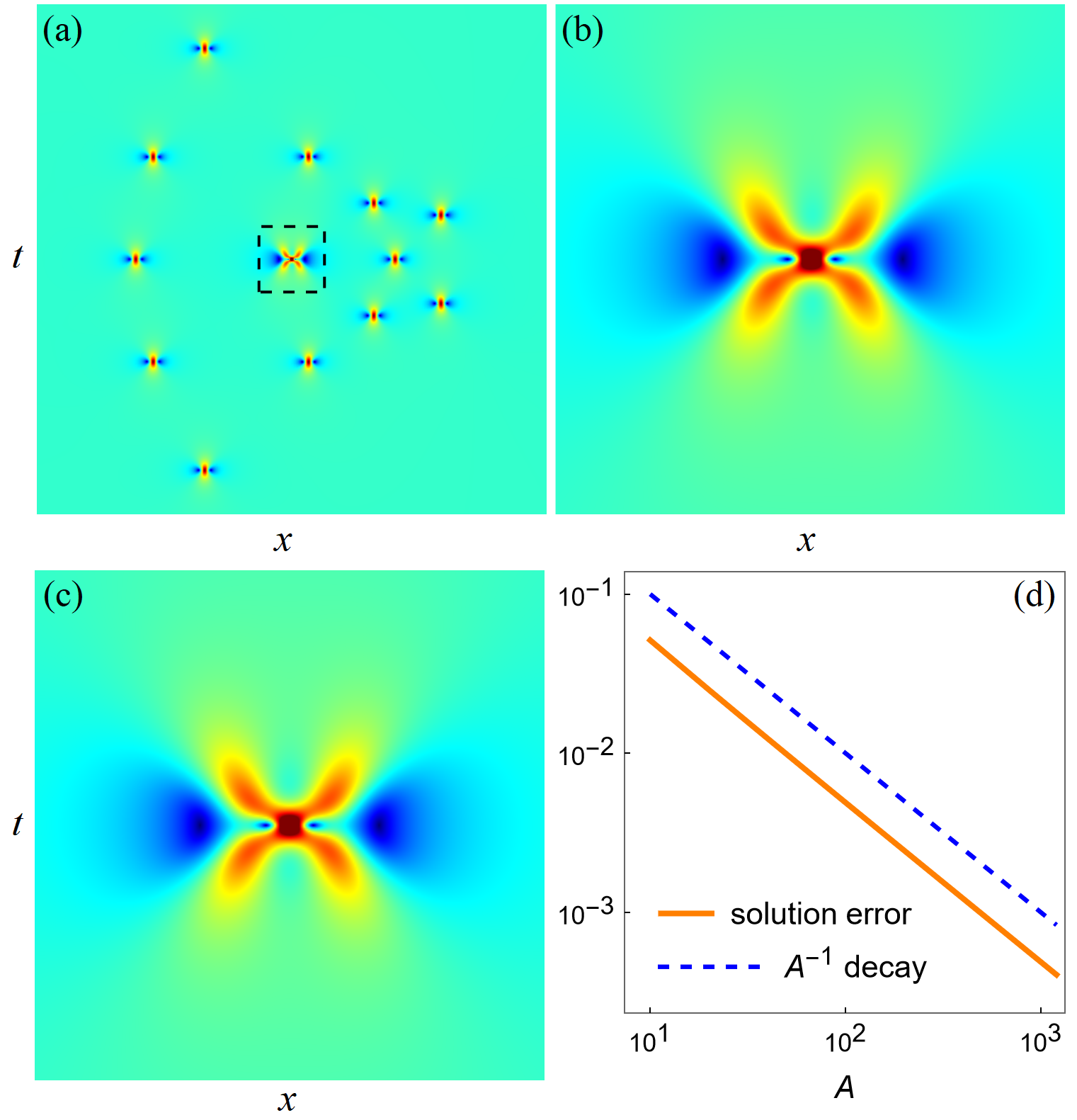}
\caption{(a) A true fifth order NLS rogue wave $|u_{5}(x, t)|$ for internal parameters (\ref{acond}) with $(\kappa_1,\kappa_2, \kappa_3, \kappa_4)=(1, 1, 1, 4/3)$ and $A=6$. (b) Zoom-in of (a) in the $(x, t)=O(1)$ region. (c) Analytical prediction of this solution in the $(x, t)=O(1)$ region from Theorem~\ref{Theorem4}. The $(x, t)$ intervals are $-35 \leq x, t  \leq 35$ for panel (a) and $-5 \leq  x,t \leq 5$ for panels (b) and (c). (d) Error of our predicted solution versus $A$ (the theoretical decay rate of $A^{-1}$ is also plotted for comparison). } \label{f:NLSz0zero}
\end{center}
\end{figure}

We would like to point out that, in special cases such as when all $\{\kappa_j\}$ are zero except for one of them, the error of
this lower-order rogue wave approximation could be smaller. For example, if $\kappa_1\ne 0$ and the other $\kappa_j$'s are zero, then when $z_0=0$ is a multiple root of $\Theta_N(z)$, the lower-order rogue wave approximation in the $(x, t)=O(1)$ region would have error of $O(a_3^{-1})=O(A^{-3})$, which is much smaller than $O(A^{-1})$ in Eq.~(\ref{uNuNhat2}). Such special cases have been reported in \cite{Yang2021a,Yangbook2024} already. But in the generic case of Theorem~\ref{Theorem4}, this error is only $O(A^{-1})$ as Fig.~\ref{f:NLSz0zero}(d) shows.

\section{Conclusion and Generalizations}

In this paper, we have studied rogue patterns associated with multiple roots of Adler--Moser polynomials under multiple large parameters (\ref{acond}) in the NLS and GDNLS equations. We first showed that the multiplicity of any multiple root in any Adler--Moser polynomial is a triangular number of the form $N_0(N_0+1)/2$ for a certain integer $N_0$. We then showed that corresponding to a nonzero multiple root of the Adler--Moser polynomial, a triangular rogue cluster would appear on the spatial-temporal plane. This triangular rogue cluster comprises $N_0(N_0+1)/2$ fundamental rogue waves forming a triangular shape, and space-time locations of fundamental rogue waves in this triangle are a linear transformation of the Yablonskii--Vorob'ev polynomial $Q_{N_0}(z)$'s root structure. In the special case where this multiple root of the Adler--Moser polynomial is zero, we showed that the associated rogue pattern is a $N_0$-th order rogue wave in the $O(1)$ neighborhood of the spatial-temporal origin. Our analytical predictions were compared to true rogue solutions and good agreement was demonstrated.
These results provide a clear and clean answer to rogue patterns induced by multiple roots of Adler--Moser polynomials under multiple large parameters (\ref{acond}).

In our derivations of the above analytical results, Lemma~1 on the row echelon form of a certain $N\times 2N$ matrix (\ref{defHN}) played a crucial role. This special matrix (\ref{defHN}) naturally appears when we attempt to investigate the multiplicity of a multiple root in an Adler--Moser polynomial and rogue patterns under multiple large parameters (\ref{acond}) in the NLS and GDNLS equations. In such investigations, the leading entries (that is, the left-most nonzero entries) of rows in the row echelon form of this matrix would give dominant or relevant contributions. The special structure of the row echelon form of this matrix given in Lemma~1 then directly leads to our main results of this paper.

Our results in this paper can be generalized in multiple directions. One direction of generalization is to other integrable equations. As is already clear from \cite{Yang2021b,Yangbook2024}, our results can readily be generalized to integrable systems whose rogue waves can be expressed as determinants featuring Schur polynomials with index jumps of two. Examples include the Boussinesq equation, the Manakov system, the three-wave resonant interaction system, the long-wave-short-wave resonant interaction system, the Ablowitz-Ladik equation, the massive Thirring model, and many others \cite{Yangbook2024}. In such systems, if multiple internal parameters in their rogue wave solutions are large as in (\ref{acond}) and the corresponding Adler--Moser polynomial admits a multiple root, then a nonzero multiple root is also expected to induce a triangular rogue cluster. If this multiple root is zero, then a lower-order rogue wave is also expected in the neighborhood of the spatial-temporal origin, except that internal parameters of this lower-order rogue wave might not be all zero, which would happen if the $k$ term in the corresponding equation (\ref{PhiPsihat}) cannot be eliminated such as in the Boussinesq equation case \cite{YangYangBoussi}.

Another direction of generalization is to multiple large internal parameters whose forms are more general than those considered in this paper. The forms of large parameters we have considered are (\ref{acond}), where each parameter contains a single power term.  For a broader class of large parameters of the dual-power form
\[ \label{ajgen}
a_{2j+1} = \kappa_j A^{2j+1}+\lambda_j A, \quad 1 \le j \le N-1,
\]
where $A$ is a large positive number and $\kappa_j, \lambda_j$ free complex constants, we can extend our analysis to this case with little modification. In this case, we can quickly show that for both the NLS equation and the GDNLS equations, if
$\tilde{\lambda}_1\ne 0$, where $\tilde{\lambda}_1\equiv \lambda_1 + [z_0+3 \textrm{i} \Im(z_0)]/6$ and $z_0$ is a multiple root of multiplicity $N_0(N_0+1)/2$ in the Adler--Moser polynomial $\Theta_N(z; \kappa_1, \dots, \kappa_{N-1})$, then a triangular rogue cluster would appear on the spatial-temporal plane. This triangular rogue cluster comprises $N_0(N_0+1)/2$ fundamental rogue waves forming a triangular shape, and space-time locations of fundamental rogue waves in this triangle are a linear transformation of the Yablonskii--Vorob'ev polynomial $Q_{N_0}(z)$'s root structure. Specifically, space-time locations $(x_0, t_0)$ of these fundamental rogue waves in the triangle are still given by Eq.~(\ref{x0t0NLS}) for the NLS equation and by Eq.~(\ref{x0t0GDNLS}) for the GDNLS equation, except that $\Omega$ in those two equations should be replaced by $\widetilde{\Omega}$, where $\widetilde{\Omega}\equiv \left(-3\tilde{\lambda}_1/4\right)^{1/3}$. Large parameters of the dual-power form (\ref{ajgen}) with special choices of $(\kappa_j, \lambda_j)$ values in the NLS equation were considered in \cite{Ling2025}. Our results above hold for general $(\kappa_j, \lambda_j)$ values as long as $\tilde{\lambda}_1\ne 0$. The case of $\tilde{\lambda}_1=0$ can also be treated through a simple extension of our analysis, and details will be omitted here.

A third direction of generalization is to the pattern analysis of rogue waves which can be expressed as determinants featuring Schur polynomials with index jumps of three under multiple large parameters. Such rogue waves appear in integrable systems such as the Manakov equations and the three-wave resonant interaction system. When a single internal parameter in such rogue waves is large, their pattern has been shown to be described by root structures of Okamoto polynomial hierarchies \cite{YangOkamoto}. The question of wave patterns under multiple large internal parameters in such rogue waves is still open. This question can be addressed through a natural extension of our analysis in this paper, and it will be pursued in the near future.

\section*{Acknowledgment}
The work of B.Y. was supported in part by the National Natural Science Foundation of China (Grant No. 12201326, 12431008).

\section*{Appendix}
\setcounter{equation}{0}
\renewcommand{\theequation}{A.\arabic{equation}}

In this appendix, we prove Lemma~1.

For the $\mathbf{H}$ matrix in Eq.~(\ref{defHN}), i.e.,
\[ \label{defHN2}
\mathbf{H}=\left(\begin{array}{ccccccccc} h_1 & 1 & & &&&    \\
                               h_3 & h_2 & h_1 & 1 &&&   \\
                               h_5 & h_4 & h_3 & h_2 & h_1 & 1 & \\
                               \vdots & \vdots & \vdots & \vdots & \vdots & \vdots & \\
                               h_{2N-1} & h_{2N-2} & h_{2N-3} &h_{2N-4} & h_{2N-5} & h_{2N-6} & \cdots & h_1 & 1
                               \end{array} \right),
\]
we denote its $j$-th column as $H_j$, where $1\le j\le 2N$. Its $H_1$ and $H_2$ vectors can be related as
\[ \label{H1H2}
\left(\begin{array}{c} h_1 \\ h_3 \\ h_5 \\ \vdots \\ h_{2N-1}\end{array}\right)
=\left(\begin{array}{cccccc} \alpha_1 &&&\\ \alpha_2 & \alpha_1 && \\ \alpha_3 & \alpha_2 & \alpha_1 & \\ \vdots & \vdots & \ddots & \ddots  \\ \alpha_N & \alpha_{N-1} & \dots & \alpha_2 & \alpha_1 \end{array}\right)\left(\begin{array}{c} 1 \\ h_2 \\ h_4 \\ \vdots \\ h_{2N-2}\end{array}\right),
\]
where $\alpha_1, \alpha_2, \dots, \alpha_N$ are constants whose values can be readily determined by sequentially solving each equation from the top down. For example, from the first equation, we get $\alpha=h_1$; from the second equation, we get $\alpha_2=h_3-\alpha_1h_2$; and so on. The system of equations (\ref{H1H2}) can be rewritten as
\[
H_1=\alpha_1H_2+\alpha_2H_4+\dots + \alpha_NH_{2N}=\sum_{j=1}^N \alpha_jH_{2j}.
\]
Since vectors $H_3, H_5, \dots$ are just zeros followed by portions of the $H_1$ vector, it is easy to see that they can be expressed through $H_4, H_6, \dots$ as well. For example,
\[
H_3=\alpha_1H_4+\alpha_2H_6+\dots + \alpha_{N-1}H_{2N}=\sum_{j=1}^{N-1} \alpha_jH_{2j+2},
\]
and so on. Using these relations, we can rewrite the $\mathbf{H}$ matrix (\ref{defHN2}) as
\[ \label{defHN3}
\mathbf{H}=\left[\sum_{j=1}^N \alpha_jH_{2j} \quad H_2 \quad \sum_{j=1}^{N-1} \alpha_jH_{2j+2} \quad H_4 \quad \dots  \quad \alpha_1H_{2N} \quad H_{2N} \right].
\]
Using the first rows of the $H_2$ vectors in this matrix and performing type-ii row operations of Sec.~\ref{secRowEch}, we can eliminate all lower rows of those $H_2$ vectors and reduce them to $[1, 0, \dots, 0]^T$, where the superscript `\emph{T}' represents the transpose of a vector. This process only affects the $H_2$ vectors; other vectors $H_4, H_6, \dots, H_{2N}$ in this $\mathbf{H}$ matrix (\ref{defHN3}) remain intact, because those other $H_{2j}$ vectors have zero as their first elements. Next, we use the second rows of the $H_4$ vectors and perform type-ii row operations of Sec.~\ref{secRowEch} to eliminate all lower rows of those $H_4$ vectors and reduce them to $[0, 1, 0,  \dots, 0]^T$. In this process, vectors $H_6, H_8, \dots, H_{2N}$ in (\ref{defHN3}) will remain intact. Continuing this process, we then find that the $\mathbf{H}$ matrix (\ref{defHN3}) can be reduced through type-ii row operations to the following matrix
\[ \label{defG}
\mathbf{G}=\left(\begin{array}{ccccccccc} \alpha_1 & 1 & & &&&    \\
                               \alpha_2 & 0 & \alpha_1 & 1 &&&   \\
                               \alpha_3 & 0 & \alpha_2 & 0 & \alpha_1 & 1 & \\
                               \alpha_4 & 0 & \alpha_3 & 0 & \alpha_2 & 0 & \cdots \\
                               \vdots & \vdots & \vdots & \vdots & \vdots & \vdots & \vdots \\
                               \alpha_{N} & 0 & \alpha_{N-1} & 0  & \alpha_{N-2} & 0 & \cdots & \alpha_1 & 1
                               \end{array} \right).
\]

Next, we further reduce this $\mathbf{G}$ matrix through type-i and type-ii row operations of Sec.~\ref{secRowEch}. Let us
denote the $j$-th column of this $\mathbf{G}$ matrix as $G_j$, where $1\le j\le 2N$. Now, we need to introduce a key parameter $k$, which is defined as the integer where
\[ \label{defk}
\mbox{Rank}(G_1, G_2, \dots, G_k)=\mbox{Rank}(G_1, G_2, \dots, G_{k+1})=k.
\]
In other words, this $k$ is the number of the first consecutive columns of $\mathbf{G}$ that are linearly independent and the addition of the next column of $\mathbf{G}$ would make them linearly dependent. Clearly, such $k$ exists and is unique, and $0\le k\le N$. In particular, when $k=0$, $\alpha_1=\alpha_2=\dots=\alpha_N=0$; and when $k=N$, $\mbox{Rank}(\mathbf{G})=N$. Since type-ii row operations do not affect the rank or linear dependence of column vectors of a matrix, this parameter $k$ can also be defined in terms of the original $\mathbf{H}$ matrix as
\[
\mbox{Rank}(H_1, H_2, \dots, H_k)=\mbox{Rank}(H_1, H_2, \dots, H_{k+1})=k.
\]
This $k$ number matches that given in Lemma~1.

The condition (\ref{defk}) means that $G_{k+1}$ is linearly dependent on $(G_1, G_2, \dots, G_k)$, i.e.,
\[  \label{Gk1cond}
G_{k+1}=\sum_{j=1}^k r_j G_j,
\]
where $r_j$'s are certain constants. This condition gives linear relations between $\alpha_j$ parameters, which we will use to reduce the matrix $\mathbf{G}$ to a row echelon form.

\vspace{0.2cm}
\noindent
\textbf{(1) The case of $r_1\ne 0$}

We first consider the case of $r_1\ne 0$ in Eq.~(\ref{Gk1cond}). In this case, we can rewrite this vector relation as
\[  \label{Gk1cond2}
G_1=\sum_{j=2}^{k+1}\hat{r}_{j}G_{j},
\]
where $\hat{r}_{j}\equiv -r_j/r_1$ for $2\le j\le k$ and $\hat{r}_{k+1}\equiv 1/r_1$.

Suppose $k$ is even, say $k=2n$ where $n$ is an integer. Since the vector $G_{k+1}$ starts with $n$ zeros and followed by $[\alpha_1, \dots, \alpha_{N-n}]^T$, the vector $G_{k}$ starts with $n-1$ zeros, followed by 1, and then $N-n$ zeros, and so on, the vector relation (\ref{Gk1cond2}) can be written out element-wise as
\begin{eqnarray}
&& \alpha_1=\hat{r}_2, \\
&& \alpha_2=\hat{r}_3\alpha_1+\hat{r}_4, \\
&& \dots\dots  \\
&& \alpha_n=\hat{r}_{3}\alpha_{n-1}+\hat{r}_5\alpha_{n-2}+\cdots+\hat{r}_{2n-1}\alpha_1+\hat{r}_{2n}, \\
&& \alpha_{n+1}=\hat{r}_{3}\alpha_{n}+\hat{r}_5\alpha_{n-1}+\cdots+\hat{r}_{2n-1}\alpha_2+\hat{r}_{2n+1}\alpha_1, \\
&& \alpha_{n+2}=\hat{r}_{3}\alpha_{n+1}+\hat{r}_5\alpha_{n}+\cdots+\hat{r}_{2n-1}\alpha_3+\hat{r}_{2n+1}\alpha_2, \\
&& \dots\dots \\
&& \alpha_{N}=\hat{r}_{3}\alpha_{N-1}+\hat{r}_5\alpha_{N-2}+\cdots+\hat{r}_{2n-1}\alpha_{N-n+1}+\hat{r}_{2n+1}\alpha_{N-n}.
\end{eqnarray}
Using these element-wise relations and performing type-ii row operations, we can reduce the $\mathbf{G}$ matrix (\ref{defG}) to the following form
\[
\widehat{\mathbf{G}}=\left(\begin{array}{cccccccccccc} \alpha_1 & 1 & & &&&    \\
                               \alpha_2 & 0 & \alpha_1 & 1 &&&   \\
                               \vdots & \vdots & \vdots & \vdots & \cdots & \cdots & \\
                               \alpha_n & 0 & \alpha_{n-1} & 0 & \alpha_{n-2} & 0 & \cdots & \cdots \\
                               0 & -\hat{r}_{2n+1} & \hat{r}_{2n} &  \cdots     \\
                               0 & 0 & 0 &  -\hat{r}_{2n+1} & \hat{r}_{2n} & \cdots     \\                                             0 & 0 & 0 & 0 & 0 & -\hat{r}_{2n+1} & \hat{r}_{2n} & \cdots     \\
                               \vdots & \vdots & \vdots & \vdots & \vdots & \vdots & \vdots & \vdots & \cdots   \\
                               0 & 0 & 0 & 0  & 0 & 0 & \cdots & 0 & -\hat{r}_{2n+1} & \hat{r}_{2n} & \cdots
                               \end{array} \right),
\]
where the first $n$ rows are unchanged from $\mathbf{G}$. The way to do it is that, we first multiply row $N-1$ of $\mathbf{G}$
by $\hat{r}_{3}$ and subtract it from row $N$, multiply row $N-2$ by $\hat{r}_5$ and subtract it from row $N$, $\dots$, and lastly multiply row $N-n$ by $\hat{r}_{2n+1}$ and subtract it from row $N$. Then, by utilizing the above explicit relations, the last row of $\mathbf{G}$ would reduce to the last row of the above $\widehat{\mathbf{G}}$ matrix, while the first $N-1$ rows of $\mathbf{G}$ remain intact. Next, we multiply row $N-2$ of $\mathbf{G}$ by $\hat{r}_{3}$ and subtract it from row $N-1$, multiply row $N-3$ by $\hat{r}_5$ and subtract it from row $N-1$, $\dots$, and lastly multiply row $N-n-1$ by $\hat{r}_{2n+1}$ and subtract it from row $N-1$. Then, by utilizing the above explicit relations again, row $N-1$ of $\mathbf{G}$ would reduce to row $N-1$ of the above $\widehat{\mathbf{G}}$ matrix, while the first $N-2$ rows of $\mathbf{G}$ remain intact. This process repeats and the above $\widehat{\mathbf{G}}$ would result from these type-ii row operations. This $\widehat{\mathbf{G}}$ matrix can be structured as
\[ \label{defGhatN2}
\widehat{\mathbf{G}}=\left(\begin{array}{cc} \widehat{\mathbf{A}} & \widehat{\mathbf{C}}    \\  \mathbf{0} & \mathbf{B}                               \end{array} \right),
\]
where $\widehat{\mathbf{A}}$ is a matrix of size $2n\times 2n$, i.e., $k\times k$, and $\mathbf{B}$ is a $(N-k)\times (2N-k)$ matrix of the form
\[ \label{Bform1}
\mathbf{B}=\left(\begin{array}{ccccccccccc}
                               0 & -\hat{r}_{2n+1} & \hat{r}_{2n}  & \cdots     \\
                               0 & 0 & 0 &  -\hat{r}_{2n+1} & \hat{r}_{2n} &   \cdots     \\                                             0 & 0 & 0 & 0 & 0 & -\hat{r}_{2n+1} & \hat{r}_{2n}  & \cdots     \\
                               \vdots & \vdots & \vdots & \vdots & \vdots & \vdots & \vdots & \vdots & \cdots \\
                               0 & 0 & 0 & 0  & 0 & 0 & \cdots & 0 & -\hat{r}_{2n+1} & \hat{r}_{2n} & \cdots
                               \end{array} \right).
\]
This form of $\mathbf{B}$ matches that in Eq.~(\ref{BformLemma1}) of Lemma~1 with $\beta=-\hat{r}_{2n+1}=-1/r_1 \ne 0$. Regarding $\widehat{\mathbf{A}}$, since $\mbox{Rank}(G_1, G_2, \dots, G_k)=k$ and type-ii row operations do not change the rank of the resulting columns, we see that the rank of the first $k$ (i.e., $2n$) columns of the $\widehat{\mathbf{G}}$ matrix is also $2n$. In view of the structure (\ref{defGhatN2}) of this $\widehat{\mathbf{G}}$ matrix, we see that the rank of the $\widehat{\mathbf{A}}$ matrix is $2n$. Thus, $\widehat{\mathbf{A}}$ is nonsingular and can be reduced to an upper triangular matrix $\mathbf{A}$ with nonzero diagonal elements through type-i and type-ii row operations. Applying these same type-i and type-ii row operations to the first $k$ rows of $\widehat{\mathbf{G}}$ in Eq.~(\ref{defGhatN2}), the resulting matrix is then the  row echelon form $\widehat{\mathbf{H}}$ of matrix $\mathbf{H}$ whose structure is as described in Lemma~1.

Next, we consider the other case where $k$ is odd, say $k=2n+1$ where $n$ is an integer. In this case, we notice that the vector $G_{k+1}$ starts with $n$ zeros, followed by 1, and then $N-n-1$ zeros; the vector $G_{k}$ starts with $n$ zeros and followed by $[\alpha_1, \dots, \alpha_{N-n}]^T$; and so on. Thus, the vector relation (\ref{Gk1cond2}) can be written out element-wise as
\begin{eqnarray}
&& \alpha_1=\hat{r}_2, \\
&& \alpha_2=\hat{r}_3\alpha_1+\hat{r}_4, \\
&& \dots\dots  \\
&& \alpha_n=\hat{r}_{3}\alpha_{n-1}+\hat{r}_5\alpha_{n-2}+\cdots+\hat{r}_{2n-1}\alpha_1+\hat{r}_{2n}, \\
&& \alpha_{n+1}=\hat{r}_{3}\alpha_{n}+\hat{r}_5\alpha_{n-1}+\cdots+\hat{r}_{2n-1}\alpha_2+\hat{r}_{2n+1}\alpha_1+\hat{r}_{2n+2}, \\
&& \alpha_{n+2}=\hat{r}_{3}\alpha_{n+1}+\hat{r}_5\alpha_{n}+\cdots+\hat{r}_{2n-1}\alpha_3+\hat{r}_{2n+1}\alpha_2, \\
&& \dots\dots \\
&& \alpha_{N}=\hat{r}_{3}\alpha_{N-1}+\hat{r}_5\alpha_{N-2}+\cdots+\hat{r}_{2n-1}\alpha_{N-n+1}+\hat{r}_{2n+1}\alpha_{N-n}.
\end{eqnarray}
Using these element-wise relations and performing type-ii row operations similar to what we did in the even-$k$ case earlier, we can reduce the $\mathbf{G}$ matrix (\ref{defG}) to the following form
\[
\widehat{\mathbf{G}}=\left(\begin{array}{cccccccccccc} \alpha_1 & 1 & & &&&    \\
                               \alpha_2 & 0 & \alpha_1 & 1 &&&   \\
                               \vdots & \vdots & \vdots & \vdots & \cdots & \cdots & \\
                               \alpha_n & 0 & \alpha_{n-1} & 0 & \alpha_{n-2} & 0 & \cdots & \cdots \\
                               \hat{r}_{2n+2} & -\hat{r}_{2n+1} &  \cdots     \\
                               0 & 0 & \hat{r}_{2n+2} &  -\hat{r}_{2n+1} & \cdots     \\                                             0 & 0 & 0 & 0 & \hat{r}_{2n+2} & -\hat{r}_{2n+1} & \cdots     \\
                               \vdots & \vdots & \vdots & \vdots & \vdots & \vdots & \vdots & \vdots & \cdots   \\
                               0 & 0 & 0 & 0  & 0 & 0 & \cdots & \hat{r}_{2n+2} & -\hat{r}_{2n+1} & \cdots
                               \end{array} \right),
\]
where the first $n$ rows are unchanged from $\mathbf{G}$. This matrix can be structured as
\[ \label{defGhatN3}
\widehat{\mathbf{G}}=\left(\begin{array}{cc} \widehat{\mathbf{A}} & \widehat{\mathbf{C}}    \\  \mathbf{0} & \mathbf{B}                               \end{array} \right),
\]
where $\widehat{\mathbf{A}}$ is a matrix of size $(2n+1)\times (2n+1)$, i.e., $k\times k$, and $\mathbf{B}$ is a $(N-k)\times (2N-k)$ matrix of the form
\[ \label{Bform2}
\mathbf{B}=\left(\begin{array}{ccccccccccc}
                               0 & \hat{r}_{2n+2} & -\hat{r}_{2n+1}  & \cdots     \\
                               0 & 0 & 0 &  \hat{r}_{2n+2} & -\hat{r}_{2n+1} &   \cdots     \\                                             0 & 0 & 0 & 0 & 0 & \hat{r}_{2n+2} & -\hat{r}_{2n+1}  & \cdots     \\
                               \vdots & \vdots & \vdots & \vdots & \vdots & \vdots & \vdots & \vdots & \cdots \\
                               0 & 0 & 0 & 0  & 0 & 0 & \cdots & 0 & \hat{r}_{2n+2} & -\hat{r}_{2n+1} & \cdots
                               \end{array} \right).
\]
This form of $\mathbf{B}$ matches that in Eq.~(\ref{BformLemma1}) of Lemma~1 with $\beta=\hat{r}_{2n+2}=1/r_1\ne 0$. Regarding $\widehat{\mathbf{A}}$, using the same arguments as for the even-$k$ case above, we see that $\widehat{\mathbf{A}}$ is nonsingular and can be reduced to an upper triangular matrix $\mathbf{A}$ with nonzero diagonal elements through type-i and type-ii row operations. Applying these same type-i and type-ii row operations to the first $k$ rows of $\widehat{\mathbf{G}}$ in Eq.~(\ref{defGhatN3}), the resulting matrix is then the row echelon form $\widehat{\mathbf{H}}$ of matrix $\mathbf{H}$ whose structure is as described in Lemma~1.

\vspace{0.2cm}
\noindent
\textbf{(2) The case of $r_1=0$ but $r_3\ne 0$}

If $r_1=0$ but $r_3\ne 0$, by examining the first equation in the vector relation (\ref{Gk1cond}), we see that $r_2=0$ as well. Thus, the vector relation (\ref{Gk1cond}) can be rewritten as
\[  \label{Gk1cond4}
G_3=\sum_{j=4}^{k+1}\hat{r}_{j}G_{j},
\]
where $\hat{r}_{j}\equiv -r_j/r_3$ for $4\le j\le k$ and $\hat{r}_{k+1}\equiv 1/r_3$.

If $k$ is even, say $k=2n$ where $n$ is an integer, then the above vector relation (\ref{Gk1cond4}) can be written out element-wise as
\begin{eqnarray}
&& 0 = 0, \\
&& \alpha_1=\hat{r}_4, \\
&& \alpha_2=\hat{r}_5\alpha_1+\hat{r}_6, \\
&& \dots\dots  \\
&& \alpha_{n-1}=\hat{r}_{5}\alpha_{n-2}+\hat{r}_7\alpha_{n-3}+\cdots+\hat{r}_{2n-1}\alpha_1+\hat{r}_{2n}, \\
&& \alpha_{n}=\hat{r}_{5}\alpha_{n-1}+\hat{r}_7\alpha_{n-2}+\cdots+\hat{r}_{2n-1}\alpha_2+\hat{r}_{2n+1}\alpha_1,  \\
&& \alpha_{n+1}=\hat{r}_{5}\alpha_{n}+\hat{r}_7\alpha_{n-1}+\cdots+\hat{r}_{2n-1}\alpha_3+\hat{r}_{2n+1}\alpha_2, \\
&& \dots\dots \\
&& \alpha_{N-1}=\hat{r}_{5}\alpha_{N-2}+\hat{r}_7\alpha_{N-3}+\cdots+\hat{r}_{2n-1}\alpha_{N-n+1}+\hat{r}_{2n+1}\alpha_{N-n}.
\end{eqnarray}
Using these element-wise relations and performing type-ii row operations similar to what we did in the $r_1\ne 0$ case earlier, we can reduce the $\mathbf{G}$ matrix (\ref{defG}) to the following form
\[
\widehat{\mathbf{G}}=\left(\begin{array}{cccccccccccc} \alpha_1 & 1 & & &&&    \\
                               \alpha_2 & 0 & \alpha_1 & 1 &&&   \\
                               \vdots & \vdots & \vdots & \vdots & \cdots & \cdots & \\
                               \alpha_{n-1} & 0 & \alpha_{n-2} & 0 & \alpha_{n-3} & 0 & \cdots & \cdots \\
                               0 & -\hat{r}_{2n+1} & \hat{r}_{2n} &  \cdots     \\
                               0 & 0 & 0 &  -\hat{r}_{2n+1} & \hat{r}_{2n} & \cdots     \\                                             0 & 0 & 0 & 0 & 0 & -\hat{r}_{2n+1} & \hat{r}_{2n} & \cdots     \\
                               \vdots & \vdots & \vdots & \vdots & \vdots & \vdots & \vdots & \vdots & \cdots   \\
                               0 & 0 & 0 & 0  & 0 & 0 & \cdots & 0 & -\hat{r}_{2n+1} & \hat{r}_{2n} & \cdots \\
                               \alpha_{N} & 0 & \alpha_{N-1} & 0  & \alpha_{N-2} & 0 & \cdots & \cdots & \cdots & \cdots & \cdots
                               \end{array} \right),
\]
where the first $n-1$ rows and the last row are unchanged from $\mathbf{G}$. Moving the last row of this $\widehat{\mathbf{G}}$ matrix above its first row (which is a type-i row operation), the resulting matrix has the structure (\ref{defGhatN2}), where $\widehat{\mathbf{A}}$ is a nonsingular matrix of size $2n\times 2n$, i.e., $k\times k$, and $\mathbf{B}$ is a $(N-k)\times (2N-k)$ matrix as given in Eq.~(\ref{Bform1}). This form of $\mathbf{B}$ matches that in Eq.~(\ref{BformLemma1}) of Lemma~1 with $\beta=-\hat{r}_{2n+1}=-1/r_3\ne 0$. The reason of the $\widehat{\mathbf{A}}$ matrix being nonsingular is the same as before, i.e., its rank is $k$ which is the same as the rank of the first $k$ columns of the $\mathbf{G}$ matrix. Since $\widehat{\mathbf{A}}$ is nonsingular, it can be reduced to an upper triangular matrix $\mathbf{A}$ with nonzero diagonal elements through type-i and type-ii row operations. Applying these same type-i and type-ii row operations to the first $k$ rows of that whole matrix, the resulting matrix is then the row echelon form $\widehat{\mathbf{H}}$ of matrix $\mathbf{H}$ whose structure is as described in Lemma~1.

If $k$ is odd, say $k=2n+1$ where $n$ is an integer, then the vector condition (\ref{Gk1cond4}) can be written out element-wise as
\begin{eqnarray}
&& 0=0, \\
&& \alpha_1=\hat{r}_4, \\
&& \alpha_2=\hat{r}_5\alpha_1+\hat{r}_6, \\
&& \dots\dots  \\
&& \alpha_n=\hat{r}_{5}\alpha_{n-1}+\hat{r}_7\alpha_{n-2}+\cdots+\hat{r}_{2n+1}\alpha_1+\hat{r}_{2n+2}, \\
&& \alpha_{n+1}=\hat{r}_{5}\alpha_{n}+\hat{r}_7\alpha_{n-1}+\cdots+\hat{r}_{2n+1}\alpha_2, \\
&& \alpha_{n+2}=\hat{r}_{5}\alpha_{n+1}+\hat{r}_7\alpha_{n}+\cdots+\hat{r}_{2n+1}\alpha_3, \\
&& \dots\dots \\
&& \alpha_{N-1}=\hat{r}_{5}\alpha_{N-2}+\hat{r}_7\alpha_{N-3}+\cdots+\hat{r}_{2n+1}\alpha_{N-n}.
\end{eqnarray}
Using these element-wise relations and performing type-ii row operations as before, we can reduce the $\mathbf{G}$ matrix (\ref{defG}) to the following form
\[
\widehat{\mathbf{G}}=\left(\begin{array}{cccccccccccc} \alpha_1 & 1 & & &&&    \\
                               \alpha_2 & 0 & \alpha_1 & 1 &&&   \\
                               \vdots & \vdots & \vdots & \vdots & \cdots & \cdots & \\
                               \alpha_{n-1} & 0 & \alpha_{n-2} & 0 & \alpha_{n-3} & 0 & \cdots & \cdots \\
                               \hat{r}_{2n+2} & -\hat{r}_{2n+1} &  \cdots     \\
                               0 & 0 & \hat{r}_{2n+2} &  -\hat{r}_{2n+1} & \cdots     \\                                             0 & 0 & 0 & 0 & \hat{r}_{2n+2} & -\hat{r}_{2n+1} & \cdots     \\
                               \vdots & \vdots & \vdots & \vdots & \vdots & \vdots & \vdots & \vdots & \cdots   \\
                               0 & 0 & 0 & 0  & 0 & 0 & \cdots & \hat{r}_{2n+2} & -\hat{r}_{2n+1} & \cdots \\
                                \alpha_{N} & 0 & \alpha_{N-1} & 0  & \alpha_{N-2} & 0 & \cdots & \cdots & \cdots & \cdots & \cdots
                               \end{array} \right),
\]
where the first $n-1$ rows and the last row are unchanged from $\mathbf{G}$. Moving the last row of this $\widehat{\mathbf{G}}$ matrix above its first row, the resulting matrix then has the structure (\ref{defGhatN3}), where $\widehat{\mathbf{A}}$ is a nonsingular matrix of size $(2n+1)\times (2n+1)$, i.e., $k\times k$, and $\mathbf{B}$ is a $(N-k)\times (2N-k)$ matrix as given in Eq.~(\ref{Bform2}). This form of $\mathbf{B}$ matches that in Eq.~(\ref{BformLemma1}) of Lemma~1 with $\beta=\hat{r}_{2n+2}=1/r_3\ne 0$. The reason of this $\widehat{\mathbf{A}}$ matrix being nonsingular is the same as before. Since $\widehat{\mathbf{A}}$ is nonsingular, it can be reduced to an upper triangular matrix $\mathbf{A}$ with nonzero diagonal elements through type-i and type-ii row operations. Applying these same type-i and type-ii row operations to the first $k$ rows of that whole matrix, the resulting matrix is then the row echelon form $\widehat{\mathbf{H}}$ of matrix $\mathbf{H}$ whose structure is as described in Lemma~1.

\textbf{(3) The remaining cases}

The above treatments can be easily extended to the remaining cases, such as $r_1=r_3=0$ but $r_5\ne 0$, $r_1=r_3=r_5=0$ but $r_7\ne 0$, and so on.

When $k$ is even where $k=2n$, the last (extreme) case is where $r_1=r_3=\cdots =r_{2n-1}=0$. In this case, Eq.~(\ref{Gk1cond}) shows that $r_2=r_4=\cdots=r_{2n}=0$ as well. Thus, $G_{k+1}=G_{2n+1}=0$, i.e., $\alpha_1=\alpha_2=\cdots=\alpha_{N-n}=0$. Because of this, the matrix $\mathbf{G}$ in Eq.~(\ref{defG}) then becomes
\[
\mathbf{G}=\left(\begin{array}{ccccccccccc}
0  & 1 &      \\
0 & 0 & 0 &  1 \\
\vdots & \vdots & \vdots  & \vdots & \cdots &\cdots  \\
0 & 0 & 0 & 0  & 0 & 0 & \cdots  & 0 & 1  \\
\alpha_{N-n+1} & 0 & 0 & 0  & 0 & 0 & \cdots & \cdots & \cdots & \cdots & \cdots \\
\alpha_{N-n+2} & 0 & \alpha_{N-n+1} & 0  & 0 & 0 & \cdots & \cdots & \cdots & \cdots & \cdots \\
\vdots & \vdots & \vdots  & \vdots & \cdots &\cdots  \\
\alpha_{N} & 0 & \alpha_{N-1} & 0  & \alpha_{N-2} & 0 & \cdots & \cdots & \cdots & \cdots & \cdots
                               \end{array}\right).
\]
Moving the last $n$ rows of this matrix above its first row, the resulting matrix is then of the form (\ref{defGhatN2}), where $\widehat{\mathbf{A}}$ is a nonsingular matrix of size $2n\times 2n$, i.e., $k\times k$, and $\mathbf{B}$ is a $(N-k)\times (2N-k)$ matrix of the form
\[ \label{Bform3}
\mathbf{B}=\left(\begin{array}{ccccccccccc}
                               0 & 1 &    \\
                               0 & 0 & 0 &  1 &      \\
                               0 & 0 & 0 & 0 & 0 & 1  &     \\
                               \vdots & \vdots & \vdots & \vdots & \vdots & \vdots & \vdots & \vdots & \cdots \\
                               0 & 0 & 0 & 0  & 0 & 0 & \cdots & 0 & 1 & \cdots
                               \end{array} \right).
\]
This form of $\mathbf{B}$ matches that in Eq.~(\ref{BformLemma1}) of Lemma~1 with $\beta=1$. The reason of the $\widehat{\mathbf{A}}$ matrix being nonsingular is the same as before. Since $\widehat{\mathbf{A}}$ is nonsingular, it can be reduced to an upper triangular matrix $\mathbf{A}$ with nonzero diagonal elements through type-i and type-ii row operations. Applying these same type-i and type-ii row operations to the first $k$ rows of that whole matrix, the resulting matrix is then the row echelon form $\widehat{\mathbf{H}}$ of matrix $\mathbf{H}$ whose structure is as described in Lemma~1.

When $k$ is odd where $k=2n+1$, the last (extreme) case is where $r_1=r_3=\cdots =r_{2n-1}=0$ but $r_{2n+1}\ne 0$ (the case of $r_{2n+1}=0$ as well cannot happen in view of Eq.~(\ref{Gk1cond})). In this extreme case, $r_2=r_4=\cdots=r_{2n}=0$. Thus, Eq.~(\ref{Gk1cond}) becomes $G_{2n+2}=r_{2n+1}G_{2n+1}$, i.e., $\alpha_1=1/r_{2n+1}\ne 0$ and $\alpha_2=\alpha_3=\cdots=\alpha_{N-n}=0$. Because of this, the matrix $\mathbf{G}$ in Eq.~(\ref{defG}) becomes
\[
\mathbf{G}=\left(\begin{array}{ccccccccccc}
\alpha_1   & 1 &      \\
0 & 0 & \alpha_1  &  1 \\
\vdots & \vdots & \vdots  & \vdots & \cdots &\cdots  \\
0 & 0 & 0 & 0  & 0 & 0 & \cdots  & \alpha_1  & 1  \\
\alpha_{N-n+1} & 0 & 0 & 0  & 0 & 0 & \cdots & \cdots & \cdots & \cdots & \cdots \\
\alpha_{N-n+2} & 0 & \alpha_{N-n+1} & 0  & 0 & 0 & \cdots & \cdots & \cdots & \cdots & \cdots \\
\vdots & \vdots & \vdots  & \vdots & \cdots &\cdots  \\
\alpha_{N} & 0 & \alpha_{N-1} & 0  & \alpha_{N-2} & 0 & \cdots & \cdots & \cdots & \cdots & \cdots
                               \end{array}\right).
\]
Moving the last $n$ rows of this matrix above its first row, then the resulting matrix has the structure (\ref{defGhatN3}), where $\widehat{\mathbf{A}}$ is a nonsingular matrix of size $(2n+1)\times (2n+1)$, i.e., $k\times k$, and $\mathbf{B}$ is a $(N-k)\times (2N-k)$ matrix of the form
\[ \label{Bform4}
\mathbf{B}=\left(\begin{array}{ccccccccccc}
                               0 & \alpha_1 & 1  & \cdots     \\
                               0 & 0 & 0 &  \alpha_1 & 1 &   \cdots     \\
                               0 & 0 & 0 & 0 & 0 & \alpha_1 & 1  & \cdots     \\
                               \vdots & \vdots & \vdots & \vdots & \vdots & \vdots & \vdots & \vdots & \cdots \\
                               0 & 0 & 0 & 0  & 0 & 0 & \cdots & 0 & \alpha_1 & 1 & \cdots
                               \end{array} \right).
\]
This form of $\mathbf{B}$ matches that in Eq.~(\ref{BformLemma1}) of Lemma~1 with $\beta=\alpha_1\ne 0$. The reason of the $\widehat{\mathbf{A}}$ matrix being nonsingular is the same as before. Since $\widehat{\mathbf{A}}$ is nonsingular, it can be reduced to an upper triangular matrix $\mathbf{A}$ with nonzero diagonal elements through type-i and type-ii row operations. Applying these same type-i and type-ii row operations to the first $k$ rows of that whole matrix, the resulting matrix is then the row echelon form $\widehat{\mathbf{H}}$ of matrix $\mathbf{H}$ whose structure is as described in Lemma~1. This completes the proof of Lemma~1.  $\Box$

\end{document}